\documentclass[iop,revtex4]{emulateapj}
\usepackage{lscape} 
\usepackage{natbib}

\newfont{\rten}{cmr10} 
\def \arcdeg{\hbox{$^\circ$}}

\begin{document}

\title{Does the presence of planets affect the frequency and properties of extrasolar Kuiper Belts? Results from the Herschel DEBRIS and DUNES surveys} 

\author{
A. Moro-Mart\'{\i}n\altaffilmark{1,2}, 
J. P. Marshall\altaffilmark{3,4,5},
G. Kennedy\altaffilmark{6},
B. Sibthorpe\altaffilmark{7}, 
B.C. Matthews\altaffilmark{8,9}, 
C. Eiroa\altaffilmark{5}, 
M.C. Wyatt\altaffilmark{6}, 
J.-F. Lestrade\altaffilmark{10}, 
J. Maldonado\altaffilmark{11}, 
D. Rodriguez\altaffilmark{12},
J.S. Greaves\altaffilmark{13}, 
B. Montesinos\altaffilmark{14}, 
A. Mora\altaffilmark{15}, 
M. Booth\altaffilmark{16}, 
G. Duch\^{e}ne\altaffilmark{17,18,19}, 
D. Wilner\altaffilmark{20}, 
J. Horner\altaffilmark{21,4}, 
}

\altaffiltext{1}{Space Telescope Science Institute, 3700 San Martin Drive Baltimore, MD 21218, USA. email: amaya@stsci.edu}
\altaffiltext{2}{Center for Astrophysical Sciences, Johns Hopkins University, Baltimore MD 21218, USA}
\altaffiltext{3}{School of Physics, University of New South Wales, Sydney, NSW 2052, Australia}
\altaffiltext{4}{Australian Centre for Astrobiology, University of New South Wales, Sydney, NSW 2052, Australia}
\altaffiltext{5}{Departamento de F\'{\i}sica Te\'orica, Universidad Aut\'onoma de Madrid, Cantoblanco, 28049, Madrid, Spain}
\altaffiltext{6}{Institute of Astronomy (IoA), University of Cambridge, Madingley Rd., Cambridge, CB3 0HA, UK}
\altaffiltext{7}{SRON Netherlands Institute for Space Research, NL-9747 AD, Groningen, Netherlands}
\altaffiltext{8}{Herzberg Astronomy and Astrophysics, National Research Council of Canada, 5071 West Saanich Road, Victoria, BC V9E 2E7, Canada}
\altaffiltext{9}{Department of Physics and Astronomy, University of Victoria, Finnerty Road, Victoria, BC, V8W 3P6, Canada}
\altaffiltext{10}{Observatoire de Paris, CNRS, 61 Av. de lÕObservatoire, F-75014, Paris, France}
\altaffiltext{11}{INAF - Osservatorio Astronomico di Palermo, Piazza Parlamento 1, I-90134 Palermo, Italy}
\altaffiltext{12}{Universidad de Chile, Camino el Observatorio 1515, Las Condes, Santiago, Chile}
\altaffiltext{13}{SUPA, School of Physics and Astronomy, University of St. Andrews, North Haugh, St. Andrews KY16 9SS, UK}
\altaffiltext{14}{Centro de Astrobiolog\'{\i}a, CSIC-INTA, ESAC Campus, P.O. Box 78, 28691 Villanueva de la Ca\~nada, Madrid, Spain}
\altaffiltext{15}{ESA-ESAC Gaia SOC. P.O. Box 78 28691 Villanueva de la Ca\~nada, Madrid, Spain}
\altaffiltext{16}{Instituto de Astrof\'{\i}sica, Pont\'{\i}ficia Universidad Cat\'olica de Chile, Vicu\~na Mackenna 4860,  7820436 Macul, Santiago, Chile}
\altaffiltext{17}{Astronomy Department, University of California, Berkeley, CA 94720, USA}
\altaffiltext{18}{Universite  Grenoble Alpes, IPAG, F-38000 Grenoble, France}
\altaffiltext{19}{CNRS, IPAG, F-38000 Grenoble, France}
\altaffiltext{20}{Harvard-Smithsonian Center for Astrophysics, 60 Garden Street, Cambridge, MA 02138, USA}
\altaffiltext{21}{Computational Engineering and Science Research Centre, University of Southern Queensland, West Street, Toowoomba Qld 4350, Australia}


\begin{abstract}
The study of the planet-debris disk connection can shed light on the formation and evolution of planetary systems and may help ``predict" the presence of planets around stars with certain disk characteristics. In preliminary analyses of subsamples of the {\it Herschel} {\it DEBRIS} and {\it DUNES} surveys, Wyatt et al. (\citeyear{2012MNRAS.424.1206W}) and Marshall et al. (\citeyear{2014A&A...565A..15M}) identified a tentative correlation between debris and the presence of low-mass planets. Here we use the cleanest possible sample out of these {\it Herschel} surveys to assess the presence of such a correlation, discarding stars without known ages, with ages $<$ 1 Gyr and with binary companions $<$100 AU to rule out possible correlations due to effects other than planet presence. In our resulting subsample of 204 FGK stars, we do not find evidence that debris disks are more common or more dusty around stars harboring high-mass or low-mass planets compared to a control sample without identified planets. There is no evidence either that the characteristic dust temperature of the debris disks around planet-bearing stars is any different from that in debris disks without identified planets, nor that debris disks are more or less common (or more or less dusty) around stars harboring multiple planets compared to single-planet systems. Diverse dynamical histories may account for the lack of correlations. The data show a correlation between the presence of high-mass planets and stellar metallicity, but no correlation between the presence of low-mass planets or debris and stellar metallicity. Comparing the observed cumulative distribution of fractional luminosity to those expected from a Gaussian distribution in logarithmic scale, we find that a distribution centered on the Solar system's value fits the data well, while one centered at 10 times this value can be rejected. This is of interest in the context of future terrestrial planet detection and characterization because it indicates that there are good prospects for finding a large number of debris disk systems (i.e. with evidence of harboring planetesimals, the building blocks of planets) with exozodiacal emission low enough to be appropriate targets for an {\it ATLAST}-type mission to search for biosignatures.

\end{abstract}


\section{Introduction}
\label{sec:into}
Planetesimals are the building blocks of planets, and mid- and far-infrared observations with {\it Spitzer} and {\it Herschel} indicate that at least 10--25\% of mature stars (ages of 10 Myr to 10 Gyr) with a wide range of masses (corresponding to spectral types A--M) harbor planetesimal disks with disk sizes of tens to hundreds AU. This frequency is a lower limit because the surveys are limited by sensitivity. The evidence for planetesimals comes from the presence of infrared emission in excess of that expected from the stellar photosphere, thought to arise from a circumstellar dust disk; because the  lifetime of the dust grains ($<$1 Myr) is much shorter than the age of the star ($>$10 Myr), it is inferred that the dust cannot be primordial but must be the result of steady or stochastic dust production generated by the collision, disruption, and/or sublimation of planetesimals (for reviews, see Wyatt \citeyear{2008ARA&A..46..339W}; Krivov \citeyear{2010RAA....10..383K}; Moro-Mart\'{\i}n \citeyear{2013pss3.book..431M}; Matthews et al. \citeyear{2014arXiv1401.0743M}).  

The Sun harbors such a debris disk, produced by the asteroids, comets, and Kuiper Belt objects (Jewitt et al. \citeyear{2009and..book...53J}) with a dust production rate that has changed significantly with time, being higher in the past when the asteroid and Kuiper belts (Kbs) were more densely populated (Booth et al. \citeyear{2009MNRAS.399..385B}). Today, the Solar system's debris disk is fainter than the faintest extrasolar debris disks we can observe with {\it Herschel} (Moro-Mart\'{\i}n \citeyear{2003AJ....125.2255M}, Vitense et al. \citeyear{2012A&A...540A..30V}), with a 3$\sigma$ detection limit at 10--20 times the level of dust in the current KB (Eiroa et al. \citeyear{2013A&A...555A..11E}; B. C. Matthews et al. 2015, in preparation). There is evidence of planetesimals around A- to M-type stars in both single- and multiple-star systems. These stars span several orders of magnitude difference in stellar luminosities, implying that planetesimal formation, a critical step in planet formation, is a robust process that can take place under a wide range of conditions. 

It is therefore not surprising that planets and debris disks coexist (Beichman et al. \citeyear{2005ApJ...622.1160B}; Moro-Mart\'{\i}n\ et al. \citeyear{2007ApJ...668.1165M},  \citeyear{2010ApJ...717.1123M}; Maldonado et al. \citeyear{2012A&A...541A..40M}; Wyatt et al. \citeyear{2012MNRAS.424.1206W}, Marshall et al. \citeyear{2014A&A...565A..15M}). However, based on {\it Spitzer} debris disk surveys, no statistical correlation has been found to date between the presence of known high-mass planets and debris disks (Moro-Mart\'{\i}n\ et al. \citeyear{2007ApJ...658.1312M}; Bryden et al. \citeyear{2009ApJ...705.1226B}; K\'osp\'al et al. \citeyear{2009ApJ...700L..73K}). These studies were focused on high-mass planets ($>$30 M$_{\oplus}$) because, at the time, the population of low-mass planets was unknown. Overall, the lack of correlation was understood within the context that the conditions to form debris disks are more easily met than the conditions to form high-mass planets, in which case one would not expect a correlation based on formation conditions; this was also consistent with the studies that showed that there is a correlation between stellar metallicity and the presence of massive planets (Santos et al. \citeyear{2004A&A...425.1013S}; Fisher \& Valenti \citeyear{2005ApJ...622.1102F}; Maldonado et al. \citeyear{2012A&A...541A..40M}), but there is no correlation between stellar metallicity and the presence of debris disks (Greaves et al. \citeyear{2006MNRAS.366..283G}; Bryden et al. \citeyear{2006ApJ...636.1098B}; Maldonado et al. \citeyear{2012A&A...541A..40M}). 

Recent results from the radial velocity surveys indicate that, similar to debris disks,  there is no correlation between the presence of low-mass planets and stellar metallicity (Ghezzi et al. \citeyear{2010ApJ...720.1290G}; Mayor et al. \citeyear{2011arXiv1109.2497M}; Buchhave et al. \citeyear{2012Natur.486..375B}). This might indicate that the conditions to form low-mass planets are more easily met than those to form high-mass planets.  A natural question to ask is whether low-mass planets and debris disks are correlated. 

A correlation between terrestrial planets in the inner region of the planetary systems and cold debris dust has been predicted to exist based on a comprehensive set of dynamical simulations consisting of high-mass planets, embryos, and inner and outer belts of planetesimals. These simulations find a strong correlation between the presence of cold dust and the occurrence of terrestrial planets because systems with cold dust imply a calm dynamical evolution where the building blocks of low-mass planets have been able to grow and survive; on the other hand, systems with dynamically active high-mass planets tend to destroy both the outer dust-producing planetesimal belt and the building blocks of the terrestrial planets (Raymond et al. \citeyear{2011A&A...530A..62R}, \citeyear{2012A&A...541A..11R}). 

{\it Herschel} observations have opened a new parameter space that allows us to explore fainter and colder debris disks, improving our knowledge of debris disk frequency, in particular around later-type stars. In addition, since the {\it Spitzer} planet-debris disk correlation studies were carried out, a large number of low-mass planets have been detected, the frequency of which can now be characterized. Tentative detection of a correlation between low-mass planets and debris disks was presented in Wyatt et al. (\citeyear{2012MNRAS.424.1206W}) from a preliminary study based on a {\it Herschel-DEBRIS} subsample of the nearest 60 G-type stars, which was also seen in the volume-limited sample of radial velocity planet host stars examined by Marshall et al. (\citeyear{2014A&A...565A..15M}).  In this paper, we revisit the planet-debris disk correlation (or lack thereof) in the {\it Herschel} {\it DEBRIS} and {\it DUNES} surveys (Matthews et al. \citeyear{2010A&A...518L.135M}; Eiroa et al. (\citeyear{2010A&A...518L.131E}), \citeyear{2013A&A...555A..11E}; B. C. Matthews et al. in preparation) to assess whether the frequency and properties of debris disks around a control sample of stars are statistically different from those around stars with high-mass or low-mass planets. In a companion paper (Marshall et al. \citeyear{2014A&A...565A..15M}), we describe the individual exoplanet host systems, their debris disks, and the disk dependencies on planetary system properties such as planet semi-major axis and eccentricity. 

The selection criteria of the different samples used in this study are presented in Section \ref{sec:sample_selec} (with a discussion of biases in Section \ref{sec:biases}). 
A detailed discussion of the statistical analysis using Kolmogorov-Smirnov (K-S), Fisher's exact, and survival analysis tests can be found in Section \ref{sec:freq_fluxratio} (regarding the frequency and properties of debris disks and their dependence on the presence of high-mass and low-mass planets), Section \ref{sec:results_met} (regarding the correlation with stellar metallicity), and Section \ref{sec:frac_lum} (regarding the distribution of the debris disk fractional luminosities). For a summary and discussion of our results the reader is directed to Section \ref{sec:summary}. 


\section{Sample selection}
\label{sec:sample_selec}

Table \ref{tab:set} lists the selection criteria of the different samples of stars used in our statistical analysis. Table \ref{tab:starinfo} gives information on their stellar parameters, and Table \ref{tab:obs1} lists the observed fluxes and photospheric estimates at 100 $\mu$m and the strength of the excess emission. Detailed information on the procedures followed in this paper for source extraction, photosphere subtraction, and SED fitting can be found in Kennedy et al. (\citeyear{2012MNRAS.426.2115K}; \citeyear{2012MNRAS.421.2264K}). 

All the stars included in this study are  drawn from the {\it Herschel} {\it DEBRIS} and {\it DUNES} surveys. {\it DEBRIS} is an unbiased volume-limited survey for M-, K-, G-, F-, and A-type stars, where the volume limits are 8.6, 15.6, 21.3, 23.6, and 45.5 pc, respectively (Phillips et al. \citeyear{2010MNRAS.403.1089P}; Matthews \citeyear{2010A&A...518L.135M}; B. C. Matthews et al. 2015, in prepraration). The {\it DUNES} survey covers mid-F- to mid-K-type stars within 20 pc (irrespective of planet or debris disk presence), plus a handful of stars within 25 pc known to harbor planets and/or debris disks (Eiroa et al. \citeyear{2010A&A...518L.131E}, \citeyear{2013A&A...555A..11E}). 

\renewcommand\thetable{1}
\begin{deluxetable*}{ll}
\tablewidth{0pc}
\tablecaption{Sample description}
\tablehead{
\colhead{Set} &
\colhead{Description}
}
\startdata
1 										& FGK stars in {\it DEBRIS} and {\it DUNES} with distances $<$20 pc, ages $>$ 100 Myr and no binary \\
                                                                                             & companions at $<$100 AU.\\
2 										& Subset from Set 1 without known planets.\\
3										& Subset from Set 1 harboring high-mass planets with masses $>$ 30 M$_{\oplus}$.\\
										& 3a: for planets at $>$ 0.1 AU.\\
										& 3b: for planets at $<$ 0.1 AU.\\
4 										& Subset from Set 1 harboring low-mass planets with masses $<$ 30 M$_{\oplus}$.\\
5										& Subset from Set 1 harboring excess emission at 100 $\mu$m, i.e. with ($F_{\rm100}-F_{\rm*,100})/\sigma_{\rm100} >$ 3).\\
6										& Subset from Set 1 with single planets.\\
7										& Subset from Set1 with multiple planets.\\
1y, 2y, 3ay, 3by, 4y, 5y 	&  		 Subsets from Sets 1--5 with ages $<$ 1 Gyr. \\
1o, 2o, 3ao, 3bo, 4o, 5o 	&  		 Subsets from Sets 1--5with ages $>$ 1 Gyr. \\
1oy, 2oy				&  		 Subsets from Sets 1 and 2 with ages 0.1--5 Gyr. \\
1oo, 2oo				&  		 Subsets from Sets 1 and 2 with ages $>$ 5 Gyr. \\
1l, 2l, 3al, 3bl, 4l, 5l 		&  		 Subsets from Sets 1--5 with metallicities smaller than the average [Fe/H] $\leqslant$ -0.12. \\
1h, 2h, 3ah, 3bh, 4h, 5h &  		 Subsets from Sets 1--5 with metallicities larger than the average [Fe/H] $>$ -0.12. \\
1t, 2t, 3at, 3bt, 4t, 5t 					&  		 Subsets from Sets 1--5 with estimated dust temperature assuming a blackbody.
\enddata
\label{tab:set}
\end{deluxetable*}

\subsection{Set 1: Control sample irrespective of planet and debris disk presence}

To maximize completeness from the {\it DEBRIS} and {\it DUNES} surveys, we selected for Set 1 all the FGK stars within 20 pc. 

The {\it Spitzer} surveys found that the upper envelope of the 70 $\mu$m debris disks emission shows a decline over the $\sim$ 100 Myr of a star's lifetime (Bryden et al. \citeyear{2006ApJ...636.1098B}; Hillenbrand et al. \citeyear{2008ApJ...677..630H}; Carpenter  \citeyear{2009ApJS..181..197C}). Therefore, to avoid introducing biases due to stellar age, we further restrict the control sample to stars with ages $>$ 100 Myr (of the stars with known ages, only three were excluded because of youth). Our stellar ages are obtained from Vican et al. (\citeyear{2012AJ....143..135V}) and Eiroa et al. (\citeyear{2013A&A...555A..11E}). Stellar ages can be very uncertain, and individual systems may end up in the wrong age bin\footnote{Comparing, for example, the stellar ages in Sierchio et al. (\citeyear{2014ApJ...785...33S}) to those in Vican et al. (\citeyear{2012AJ....143..135V}),  among the 48 stars that these two studies have in common, we find that differences in ages are less than 50\%, except for five stars: HD126660/HIP70497 (80\%), HD23754/HIP17651(83\%), HD189245/HIP98470 (733\%), HD20630/HIP15457 (70\%) and HD101501/HIP56997 (84\%). The age estimations are therefore broadly consistent}. However, for a statistical analysis such as the one in this paper, the best approach is to use an age database as "uniform" as possible. Our ages are based on gyrochronology, Ca$_{II}$ chromospheric emission (R'$_{HK}$) and X-ray flux, always in that order of priority, acknowledging the decreasing reliability of the corresponding age measurements. Gyrochronology ages come from Vican et al. (\citeyear{2012AJ....143..135V}) and are available for 17 stars in our sample; they can be unreliable for young stars ($<$ 300 Myr), but out of those 17 stars, only one star is in that age range. When several chromospheric ages are available, we favored the ages in Eiroa et al. (\citeyear{2013A&A...555A..11E}) over those in Vican et al. (\citeyear{2012AJ....143..135V}) because the latter were based on a literature search, whereas the former were derived using spectra obtained by the DUNES team and their innerly consistent estimates of Ca$_{II}$ activity index (out of the 162 chromospheric ages used, 107 come from Eiroa et al. \citeyear{2013A&A...555A..11E}). Stars without estimated ages were excluded from our analysis. 

We do not include A-type stars in this study because the planet searches around these targets are preferentially done around evolved A-type stars (classes III, III-IV, and IV) with lower jitter and narrower absorption lines (Johnson et al. \citeyear{2011AJ....141...16J}), whereas the A-type stars targeted by {\it DEBRIS} are main-sequence (class V). Therefore, we do not have information on planet presence for most A-type stars in the {\it DEBRIS} survey. Regarding M-type stars, 89 were observed by {\it DEBRIS}, three harboring planets, one of which also harbors a debris disk (GJ 581 - Lestrade et al. \citeyear{2012A&A...548A..86L}). We do not include M-type stars in this study because of low number statistics and because they might probe into a different regime of planetesimal and planet formation than the FGK-type stars. 

The {\it DEBRIS} and {\it DUNES} surveys include single and binary/multiple stars. Previous studies indicate that there are differences in both disk frequency and planet frequency between singles and binaries, and these could introduce a bias in our statistical analysis.  Regarding disk frequency,  Rodriguez \& Zuckerman (\citeyear{2012ApJ...745..147R}) found that, out of a sample of 112 main-sequence debris disk stars, 25\%$\pm$4\% were binaries, significantly lower than the expected 50\% for field stars, with a lack of binary systems at separations of 1--100 AU; for the debris disk hosts in the {\it DEBRIS} sample, the multiplicity frequency is $\sim$ 28\% (D .R .Rodriguez et al. 2015, in preparation). Regarding planet frequency, Eggenberger et al. (\citeyear{2007A&A...474..273E}; \citeyear{2011IAUS..276..409E}) carried out a survey with VLT/NACO to look for stellar companions around 130 nearby solar-type stars and found that the difference in binarity fraction between the nonplanet hosts and the planet hosts is 13.2\%$\pm$5.1\% for binary separations $<$ 100 AU.  In a more recent study, Wang et al. (\citeyear{2014ApJ...783....4W}) compared the stellar multiplicity of field stars to that of a sample of 138 bright Kepler multiplanet candidate systems, finding also that, for the planet hosts, the binary fraction is significantly lower than field stars for binary semimajor axes $<$20 AU. An additional observation is that, even within the giant planet regime, binaries tighter than 100 AU show a different distribution of masses, suggesting a different formation mechanism and/or dynamical history (Duchene \citeyear{2010ApJ...709L.114D}). In view of all these studies, we have excluded from our samples 96 binary systems with semi-major axis $<$100 AU to avoid introducing a bias in our analysis. In doing that, we are naturally excluding all circumbinary disks (Kennedy et al. \citeyear{2012MNRAS.421.2264K}; D. R. Rodriguez et al. 2015, in preparation), limiting our analysis to those that are circumstellar. This seems appropriate because one would expect that the degree to which the dust is affected by planets (if present) is different, whether the dust is circumbinary or circumstellar, and this could again bias any potential planet-disk correlation. 

Differences in infrared background levels could introduce a bias to the debris disk detection; however,  both the {\it DUNES} and {\it DEBRIS} surveys excluded targets that were predicted to be in regions with high contamination from galactic cirrus\footnote{The unconfirmed planethost star $\alpha$ Cen B was observed as part of the {\it DUNES} and {\it Hi-Gal} programs, but it was excluded from this analysis because its high background level does not  fulfill the {\it DUNES} and {\it DEBRIS} selection criteria, and our analysis is intended to be unbiased.}.  In addition, all the targets in Set 1 have been inspected to exclude, to the best of our knowledge, sources subject to confusion. 

The total number of stars in Set 1 (FGK stars within 20 pc, ages $>$ 100 Myr and no binary companions at $<$100 AU) is 204. All the other star samples discussed in the subsections below are extracted from Set 1, i.e. they fulfill the same criteria with respect to stellar type, distance, age, absence of close binary companions, and nearby confusion. 

Table \ref{tab:planetprop} lists the planetary systems found within Set 1.  There are 22 stars harboring planets and an additional three with unconfirmed planetary systems, namely HD 22049 ($\epsilon$ Eri), HD 10700 ($\tau$ Cet) and HD 189567.

\begin{figure}
\begin{center}
\includegraphics[width=7cm]{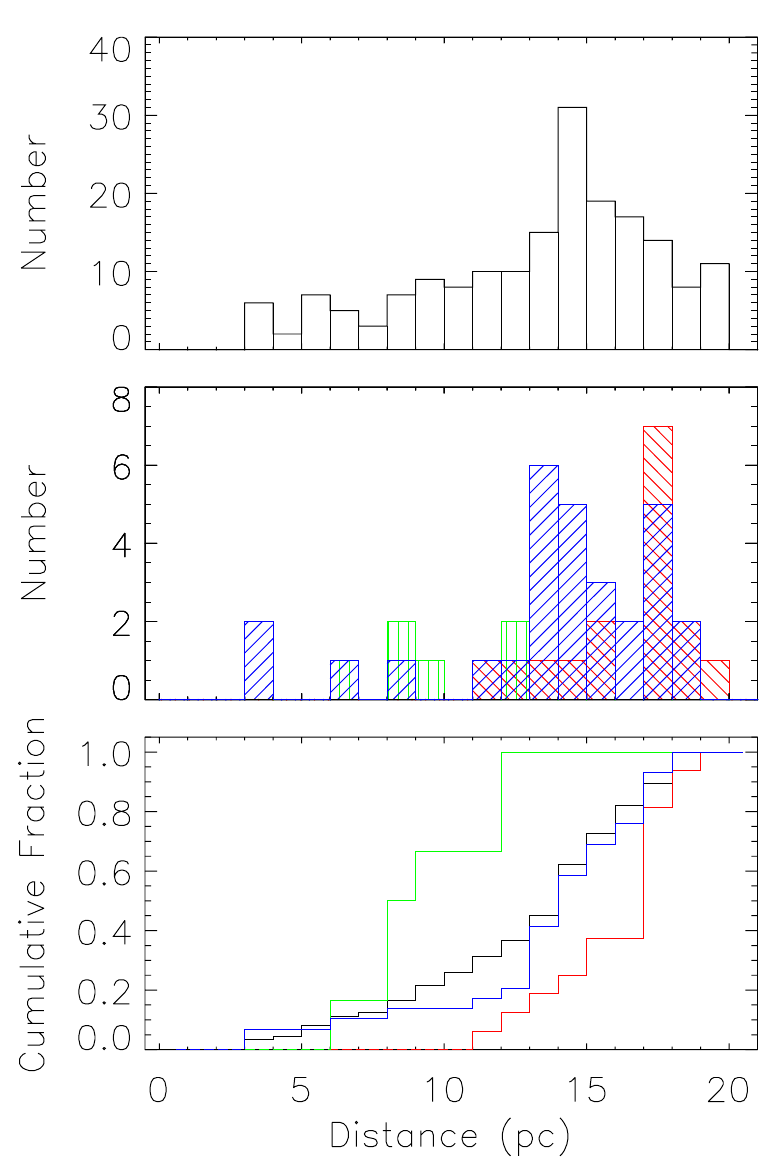}
\end{center}
\caption{Distribution of distances. {\bf Top}: Stars without known planets (Set 2). {\bf Middle}: The line-filled colored histograms  correspond to the  high-mass planet sample (Set 3; in red, with hatching from the top-left to the bottom right), low-mass planet sample (Set 4; in green, with vertical hatching) and debris disk sample (Set 5; in blue, with hatching from the top-right to the bottom left). {\bf Bottom}: Cumulative fraction of distances (same color code as above). 
}
\label{dist_100Myr_histo}
\end{figure}

Even though the targets are located at a range of distances (see Figure \ref{dist_100Myr_histo}), we do not expect this to introduce a significant bias to the planet-debris disk correlation study presented in this paper for the following reasons. Regarding planet detection, the Doppler studies do not depend on distance (although their sensitivity depends on V magnitude and spectral type, and this may account for the closer distances of stars hosting low-mass planets only). Regarding debris disk detection, (a) the {\it DUNES} observations are designed to always reach the stellar photosphere at 100 $\mu$m to a uniform signal-to-noise ratio $>$ 5; (b) we assess the planet-debris correlation using survival analysis that takes into account the upper limits  from the {\it DEBRIS} survey; and (c) we use a distance-independent variable, the dust excess flux ratio ($F_{\rm obs}^{\rm100}-F_{\rm star}^{\rm100})/F_{\rm star}^{\rm100}$, where $F_{\rm obs}^{\rm100}$ is the observed flux at 100 $\mu$m  and $F_{\rm star}^{\rm100}$  is the expected photospheric value at that wavelength. 

\subsection{Set 2: No-planet sample}

Set 2 is the subset of stars from Set 1 without known planets, as of August 2014. The number of stars in this set is 182 (179 if including the three unconfirmed planetary systems).

\subsection{Set 3: High-mass planet sample}

Set 3 is the subset of stars from Set 1  known as of August 2014 to harbor one or more planets with masses $>$ 30 M$_{\oplus}$ ($>$ 0.094 M$_{Jup}$). We call this the high-mass planet sample. The planetary system properties are listed in Table \ref{tab:planetprop}. The number of stars in this set is 16 (17 if including the three unconfirmed planetary systems). Note that some of these systems also harbor low-mass planets. We chose this limiting planet mass because for stars harboring planets $>$ 30 M$_{\oplus}$, there is a correlation between the presence of planets and stellar metallicity (Santos et al. \citeyear{2004A&A...425.1013S}; Fisher \& Valenti \citeyear{2005ApJ...622.1102F}). On the other hand, for stars harboring planets  $<$ 30 M$_{\oplus}$, there is no correlation between the presence of planets and stellar metallicity (Ghezzi et al. \citeyear{2010ApJ...720.1290G}; Mayor et al. \citeyear{2011arXiv1109.2497M}). This might indicate differences in the planet formation mechanism that may affect the planet-debris disk correlation.  We further divide this set into two subsets: 3a (for planets with $\it{a} >$ 0.1 AU) and  3b (for planets with $\it{a} <$ 0.1 AU).

\subsection{Set 4: Low-mass planet sample}

Set 4 is the subset of stars from Set 1  known as of August 2014 to harbor one or more planets with masses $<$ 30 M$_{\oplus}$ and no higher-mass planets. We call this the low-mass planet sample. There are six stars in this set (eight if including the three unconfirmed planetary systems).   

\subsection{Set 5: Debris disk sample}

Due to the wavelength coverage of the {\it DUNES} and {\it DEBRIS} surveys\footnote{{\it DEBRIS} and {\it DUNES} utilized the simultaneous 100 $\mu$m and 160 $\mu$m imaging mode as the basis for their survey data, with both teams taking additional data toward selected sources using the 70 $\mu$m and 160 $\mu$m imaging mode of PACS and 250 $\mu$m, 350 $\mu$m and 500 $\mu$m imaging with SPIRE as appropriate.}, this study is focused on the 100 $\mu$m emission. Set 5 is the subset of 29 stars from Set 1 with debris disks detected by {\it Herschel} at 100 $\mu$m, i.e. stars for which the signal to noise ratio of the excess emission is SNR$_{\rm dust} > 3$, where SNR$_{\rm dust} = {F_{\rm obs}^{\rm100}-F_{\rm star}^{\rm100} \over \sqrt{{\sigma_{\rm obs}^{\rm100}}^2+{\sigma_{\rm star}^{\rm100}}^2}}$, and $F_{\rm obs}^{\rm100}$ and $F_{\rm star}^{\rm100}$ are the observed flux at 100 $\mu$m and the estimated photospheric flux, respectively, whereas $\sigma_{\rm obs}^{\rm100}$ and $\sigma_{\rm star}^{\rm100}$ are their 1-$\sigma$ uncertainties. The 70 $\mu$m {\it Spitzer} observations do not identify any additional debris disks within Set 1. This indicates that the 100 $\mu$m emission is a good tracer of the cold KB-like dust, and we will use it as our reference wavelength. The analysis presented in this paper is limited to cold KB-like debris disks (where cold refers to  debris disks detected at 70--100 $\mu$m); we are not including the warm debris disks identified by {\it Spitzer} at 24 $\mu$m and with no excess at 70 $\mu$m (under this category there is only one planet-bearing star, HD 69830). 

Note that there are several targets harboring debris disks and/or planets that were observed with {\it Spitzer} but were not observed by the {\it Herschel} {\it DEBRIS} and {\it DUNES} surveys because of their high level of background emission. 
 
\subsection{Sets 6 and 7: Single-/Multiple-planet sample}

Set 6 is the subset of stars from Set 1 known as of August 2014 to harbor single-planet systems, while Set 7 is the subset of stars with multiple known planets. 

\subsection{Sets 1y--5y and 1o--5o: Young/Old samples}

\begin{figure}
\begin{center}
\includegraphics[width=7cm]{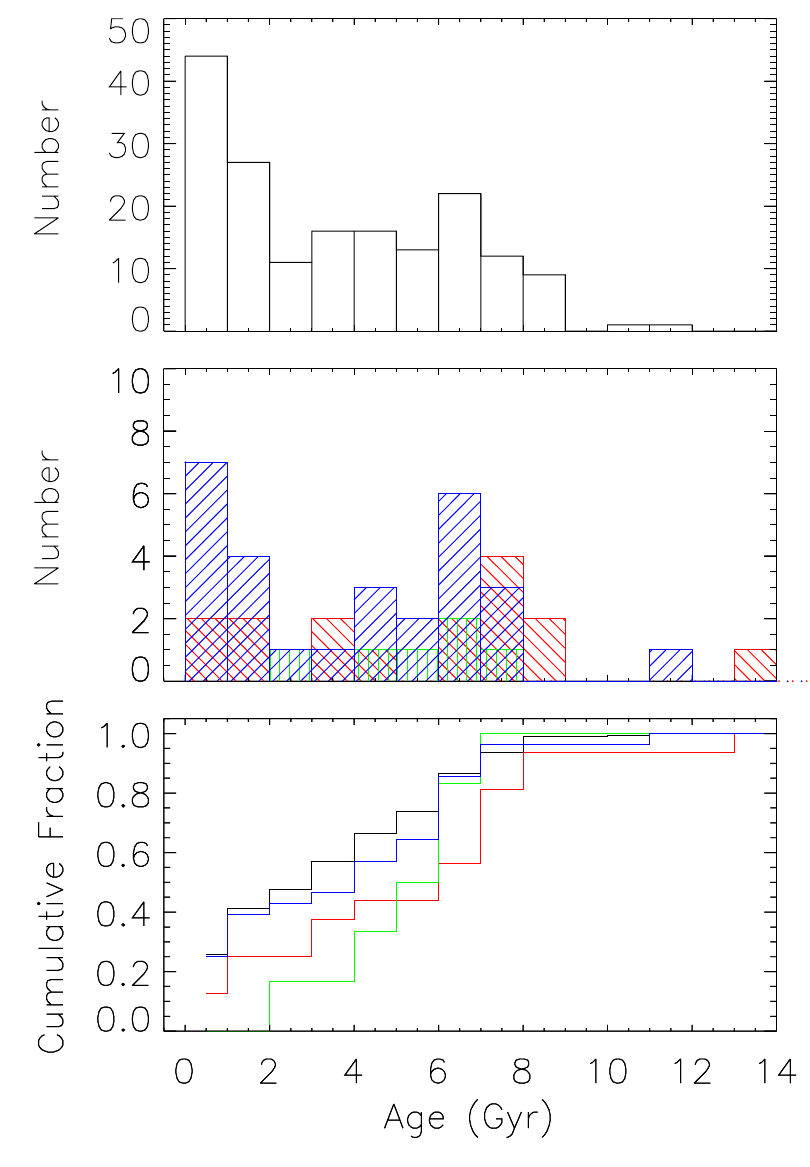}
\end{center}
\caption{Distribution of stellar ages. {\bf Top}: Stars without known planets (Set 2). {\bf Middle}: The line-filled colored histograms  correspond to the  high-mass planet sample (Set 3; in red, with hatching from the top-left to the bottom right), low-mass planet sample (Set 4; in green, with vertical hatching) and debris disk sample (Set 5; in blue, with hatching from the top-right to the bottom left). {\bf Bottom}: Cumulative distribution of stellar ages (same color code as above). 
}
\label{age_100Myr_histo}
\end{figure}

If debris disks evolve with time, and the samples compared have different age distributions, this will introduce a bias in our analysis. We therefore divide the samples into stars younger than 1 Gyr (labeled as Sets 1y--5y) and stars older than 1 Gyr (Sets 1o--5o; our sample has no hot Jupiters in Set 3o), limiting the comparison to sets of similar ages (i.e., within the o or y groups). We find that the distribution of ages in the samples considered (Figure \ref{age_100Myr_histo}) show that planet-bearing stars (Sets 3 and 4) tend to be older on average than the stars in the no-planet sample (Set 2); this is because Gyr-old stars have low magnetic activity, implying lower levels of radial velocity jitter that facilitate the Doppler studies. While this might result in planet-bearing stars having fewer debris detections if debris levels decrease with age, Figure \ref{age_100Myr_histo} shows little evidence for evolution in disk detectability with time, and this is discussed further in section \ref{sec:age}.

\subsection{Sets 1h and 1l: High/Low metallicity samples}

To explore the role of stellar metallicity, we divide Set 1 into two subsamples, a high-metallicity sample (Set 1h) and a low-metallicity sample (Set 1l),  using the midpoint of the metallicity distribution of Set 1, [Fe/H] = -0.12, as the dividing value. 

\section {Debris disk frequency and dust flux ratio}
\label{sec:freq_fluxratio}

The observed debris disk frequencies are listed in Tables \ref{tab:disk_freq}, \ref{tab:met} and \ref{tab:sptype}. Due to the small sample size, the statistical uncertainties are calculated using a binomial distribution rather than the $\sqrt N$ Poisson uncertainty (see the appendix of Burgasser et al. \citeyear{2003ApJ...586..512B}). Table \ref{tab:disk_freq} shows that the control sample (Set 1) has a debris disk frequency of 0.14$^{-0.02}_{+0.03}$, similar to that found by the {\it Spitzer} surveys at 70 $\mu$m (Trilling et al. \citeyear{2008ApJ...674.1086T}; Hillenbrand et al. \citeyear{2008ApJ...677..630H}; Carpenter et al. \citeyear{2009ApJS..181..197C}). This result is also in agreement with Gaspar et al. (\citeyear{2013ApJ...768...25G}) who found a Spitzer incidence rate of 17.5\% within the DUNES sample. 

\renewcommand\thetable{5}
\begin{deluxetable*}{lccccc}
\tablewidth{0pc}
\tablecaption{Debris disk frequency (at 100 $\mu$m)}
\tablehead{
\colhead{} & {} & \multicolumn{2}{c}{Excluding unconfirmed planets\tablenotemark{a}} & \multicolumn{2}{c}{Including unconfirmed planets\tablenotemark{a}} \\
\colhead{} & 
\colhead{Set} & 
\colhead{${\rm No.~of~excesses \over \rm No.~of~stars}$} &
\colhead{Excess freq.\tablenotemark{b}} & 
\colhead{${\rm No.~of~excesses \over \rm No.~of~stars}$} &
\colhead{Excess freq.\tablenotemark{b}}\\
\colhead{} &
\colhead{} &
\colhead{} &
\colhead{(at 100 $\mu$m)} &
\colhead{} &
\colhead{(at 100 $\mu$m)}
}
\startdata
$_{1}$   & 1 			&  29/204 & 0.14$^{-0.02}_{+0.03}$  & 29/204 & 0.14$^{-0.02}_{+0.03}$ \\
$_{2}$   & 2	 		&  24/182 & 0.13$^{-0.02}_{+0.03}$  & 22/179 & 0.12$^{-0.02}_{+0.03}$ \\
$_{3}$   & 3a,b 		&  3/16	  & 0.19$^{-0.06}_{+0.13}$  & 4/17     & 0.23$^{-0.07}_{+0.13}$ \\
$_{4}$   & 4			&  2/6	  & 0.33$^{-0.13}_{+0.21}$  & 3/8	  & 0.37$^{-0.13}_{+0.18}$ \\
$_{5}$   & 5			& 29/29	  & 		\nodata	    & 29/29	  & 		\nodata	      \\
$_{6}$   & 6			& 3/12	  & 0.25$^{-0.08}_{+0.15}$  & 4/14	  & 0.29$^{-0.09}_{+0.14}$ \\
$_{7}$   & 7			& 2/10	  & 0.20$^{-0.07}_{+0.17}$  & 3/11	  & 0.27$^{-0.09}_{+0.16}$ \\
\\
\hline
\\
$_{8}$   & 1y		       & 7/48	& 0.15$^{-0.04}_{+0.07}$ & 7/48	& 0.15$^{-0.04}_{+0.07}$\\
$_{9}$   & 2y			& 7/46	& 0.15$^{-0.04}_{+0.07}$ & 7/46	& 0.15$^{-0.04}_{+0.07}$ \\
$_{10}$   & 3aby		& 0/2	       & 0 			    & 0/2	&	0 \\
$_{11}$   & 4y			& 0/0		& 			    & 0/0	& 	\\		
$_{12}$ & 5y			& 7/7		& \nodata			    & 7/7  &    \nodata  \\		
\\
\hline
\\
$_{13}$ & 1o			& 21/146   & 0.14$^{-0.02}_{+0.03}$ & 21/146 & 0.14$^{-0.02}_{+0.03}$ \\
$_{14}$ & 2o	          	& 16/126   & 0.13$^{-0.02}_{+0.03}$ & 14/123 & 0.11$^{-0.02}_{+0.04}$ \\
$_{15}$ & 3abo 		       & 3/14       & 0.21$^{-0.07}_{+0.14}$ & 4/15     & 0.27$^{-0.08}_{+0.14}$ \\ 	
$_{16}$ & 4o			& 2/6	   & 0.33$^{-0.13}_{+0.21}$ & 3/8	   & 0.37$^{-0.13}_{+0.18}$ \\	
$_{17}$ & 5o			& 21/21     & 	\nodata		   & 21/21   & \nodata\\
$_{18}$ & 6o			& 3/10	  & 0.30$^{-0.10}_{+0.17}$  & 4/12	  & 0.33$^{-0.10}_{+0.15}$ \\
$_{19}$ & 7o			& 2/10	  & 0.20$^{-0.07}_{+0.17}$  & 3/11	  & 0.27$^{-0.09}_{+0.16}$ 
\enddata
\tablenotetext{a}{Unconfirmed planetary systems are HD 22049 ($\epsilon$ Eri), HD 10700 ($\tau$ Cet) and HD 189567.}
\tablenotetext{b}{The statistical uncertainties are calculated using a binomial distribution.} 
\label{tab:disk_freq}
\end{deluxetable*}

\renewcommand\thetable{6}
\begin{deluxetable*}{ccccc}
\tablewidth{0pc}
\tablecaption{Dependence with stellar metallicity}
\tablehead{
\colhead{Set} & 
\colhead{No. of stars} & 
\colhead{No. with} &
\colhead{No. with} &
\colhead{No. with}\\
\colhead{} & 
\colhead{in set} & 
\colhead{high-mass planets\tablenotemark{a}} &
\colhead{low-mass planets\tablenotemark{a}} &
\colhead{debris disks}\\
\colhead{} & 
\colhead{} & 
\colhead{($>$ 30 M$_\oplus$)} &
\colhead{($<$ 30 M$_\oplus$)} &
\colhead{(at 100 $\mu$m)}
}
\startdata
1l ([Fe/H] $\leqslant$ -0.12) 		& 61			& 1					& 3 (5)				& 9\\			
1h	([Fe/H] $>$ -0.12) 		& 75			& 14 (15)				& 3					& 17
\enddata
\tablenotetext{a}{Excluding unconfirmed planetary systems around HD 22049 ($\epsilon$ Eri), HD 10700 ($\tau$ Cet) and HD 189567. The parenthesis shows the result when including these three planetary systems.}
\label{tab:met}
\end{deluxetable*}

\renewcommand\thetable{7}
\begin{deluxetable*}{lccccccccc}
\tablewidth{0pc}
\tablecaption{Debris disk frequency (at 100 $\mu$m) as a function of spectral type}
\tablehead{
\colhead{} & {} & \multicolumn{2}{c}{Total\tablenotemark{a}} & \multicolumn{2}{c}{F-type\tablenotemark{a}} & \multicolumn{2}{c}{G-type\tablenotemark{a}} & \multicolumn{2}{c}{K-type\tablenotemark{a}} \\
\colhead{} & 
\colhead{Set} & 
\colhead{${\rm No.~of~excesses \over \rm No.~of~stars}$} &
\colhead{Excess freq.\tablenotemark{b}} & 
\colhead{${\rm No.~of~excesses \over \rm No.~of~stars}$} &
\colhead{Excess freq.\tablenotemark{b}} &
\colhead{${\rm No.~of~excesses \over \rm No.~of~stars}$} &
\colhead{Excess freq.\tablenotemark{b}} & 
\colhead{${\rm No.~of~excesses \over \rm No.~of~stars}$} &
\colhead{Excess freq.\tablenotemark{b}}\\
\colhead{} &
\colhead{} &
\colhead{} &
\colhead{(at 100 $\mu$m)} &
\colhead{} &
\colhead{(at 100 $\mu$m)} &
\colhead{} &
\colhead{(at 100 $\mu$m)} &
\colhead{} &
\colhead{(at 100 $\mu$m)}
}

\startdata
$_{1}$   & 1 			&  29/204 & 0.14$^{-0.02}_{+0.03}$  & 10/46 & 0.22$^{-0.05}_{+0.07}$	& 11/61 & 0.18$^{-0.04}_{+0.06}$ 	       & 8/97 & 0.08$^{-0.02}_{+0.04}$ \\
$_{2}$   & 2	 		&  24/182 & 0.13$^{-0.02}_{+0.03}$  & 9/42	 & 0.21$^{-0.12}_{+0.08}$	& 7/48   & 0.15$^{-0.04}_{+0.06}$		& 8/92 & 0.09$^{-0.02}_{+0.04}$ \\
$_{3}$   & 3a,b 		&  3/16	  & 0.19$^{-0.06}_{+0.13}$  & 1/4	 & 0.25$^{-0.10}_{+0.25}$	& 2/9     & 0.22$^{-0.08}_{+0.18}$		& 0/3	    & 0\\
$_{4}$   & 4			&  2/6	  & 0.33$^{-0.13}_{+0.21}$  & 0/0	 & 				              & 2/4	& 0.5$^{-0.2}_{+0.2}$		       & 0/2	     & 0 \\
$_{5}$   & 5			& 29/29	  & 		\nodata	                & 10/10 & 	\nodata			& 11/11 & 	\nodata				& 8/8 & \nodata \\
\\
\hline
\\
$_{6}$ & 1o			& 21/146   & 0.14$^{-0.02}_{+0.03}$ & 8/33 & 0.24$^{-0.06}_{+0.09}$	& 7/49 & 0.14$^{-0.04}_{+0.06}$		& 6/64 & 0.09$^{-0.02}_{+0.05}$ \\
$_{7}$ & 2o	          	& 16/126   & 0.13$^{-0.02}_{+0.03}$ & 7/30 & 0.23$^{-0.06}_{+0.09}$	& 3/37 & 0.08$^{-0.03}_{+0.06}$		& 6/59 & 0.10$^{-0.04}_{+0.03}$ \\
$_{8}$ & 3abo 		& 3/14       & 0.21$^{-0.07}_{+0.14}$ & 1/3    & 0.33$^{-0.14}_{+0.29}$	& 2/8   & 0.25$^{-0.09}_{+0.19}$	 	& 0/3	    & 0	\\
$_{9}$ & 4o			& 2/6	   & 0.33$^{-0.13}_{+0.21}$ & 0/0    & 				              & 2/4   & 0.5$^{-0.2}_{+0.2}$		       & 0/2	     & 0 \\
$_{10}$ & 5o		& 21/21     & 	\nodata		                & 8/8    & 	\nodata			& 7/7     & 		\nodata			& 6/6	     & \nodata
\enddata
\tablenotetext{a}{Excluding unconfirmed planetary systems around HD 22049 ($\epsilon$ Eri), HD 10700 ($\tau$ Cet) and HD 189567.}
\tablenotetext{b}{The statistical uncertainties are calculated using a binomial distribution.}
\label{tab:sptype}
\end{deluxetable*}

\subsection {Dependence on stellar age}
\label{sec:age}

If debris disks evolve with time, and the samples compared have different age distributions within the decay timescale, this will introduce a bias in the comparison of the debris disk frequencies and dust flux ratios. As mentioned above, Figure \ref{age_100Myr_histo} indicates that planet-bearing stars (Sets 3 and 4) tend to be older on average than the stars in the control samples because they are preferentially targeted by the Doppler studies.   

To test for disk evolution, we divide the samples into stars with ages 0.1--1 Gyr (labeled as Sets 1y--5y) and stars older than 1 Gyr (Sets 1o--5o). We then compare the disk frequencies and dust flux ratios in the young and old samples, Set 2y and 2o (lines 9 and 14 in Table \ref{tab:disk_freq}). We do this exercise in the no-planet sample to minimize the effect of planet presence, as the goal is to check for disk evolution alone. Comparing Set 2y (with a disk frequency of 7/46 = 0.15) and Set 2o (with a disk frequency of 16/126 = 0.13) and using a binomial distribution, we find that detecting seven or more disks in Set 2y, when the expected detection rate is 0.13 (taking Set 2o as reference, i.e. the expected number of disk detections is 0.13$\cdot$46) is a 39\% probability event (24\% if including the unconfirmed planetary systems -- Table \ref{tab:stat_result}, lines 1 and 2). This probability is not low enough to claim that the higher incidence rate in the young sample compared to the old sample is significant.

The latter, however, does not take into account the uncertainty in the expected rate of the reference sample (in this case, Set 2o). The Fisher exact test is more appropriate in this regard. To carry out this test, we classify the stars in the two samples in two categories regarding disk presence: stars with disks (SNR$_{\rm dust} > 3$) and without disks (SNR$_{\rm dust} < 3$). The null hypothesis in this case is that both sets (2y and 2o) are equally likely to harbor disks. The test gives a 60\% probability to find the observed arrangement of the data if the null hypothesis were true (Table \ref{tab:stat_result}, lines 3 and 4). Note that the Fisher exact test can only reject the null hypothesis, never to prove it true. The Fisher exact test in this case does not identify any evolution in disk frequency within the timescale considered. 

A variable that is commonly used to characterize the strength of the disk emission is the dust flux ratio, ($F_{\rm obs}^{\rm100}-F_{\rm star}^{\rm100})/F_{\rm star}^{\rm100}$, where $F_{\rm obs}^{\rm100}$ is the observed flux at 100 $\mu$m  and $F_{\rm star}^{\rm100}$  is the expected photospheric value at that wavelength.  Table \ref{tab:obs1} lists the observed dust flux ratio for all the stars in our study. The 3$\sigma$ upper limits (preceded by "$<$" symbol) are given for stars without significant detected emission and are calculated assuming the observed flux is $F_{\rm obs}^{\rm100}+3\sigma_{\rm obs}^{\rm100}$, for stars with $0 < F_{\rm obs}^{\rm100}/\sigma_{\rm obs}^{\rm100} < 3$, and 3$\sigma_{\rm obs}^{\rm100}$, for stars with $F_{\rm obs}^{\rm100}/\sigma_{\rm obs}^{\rm100} < 0$. 

\begin{figure}
\centering
\includegraphics[width=7cm]{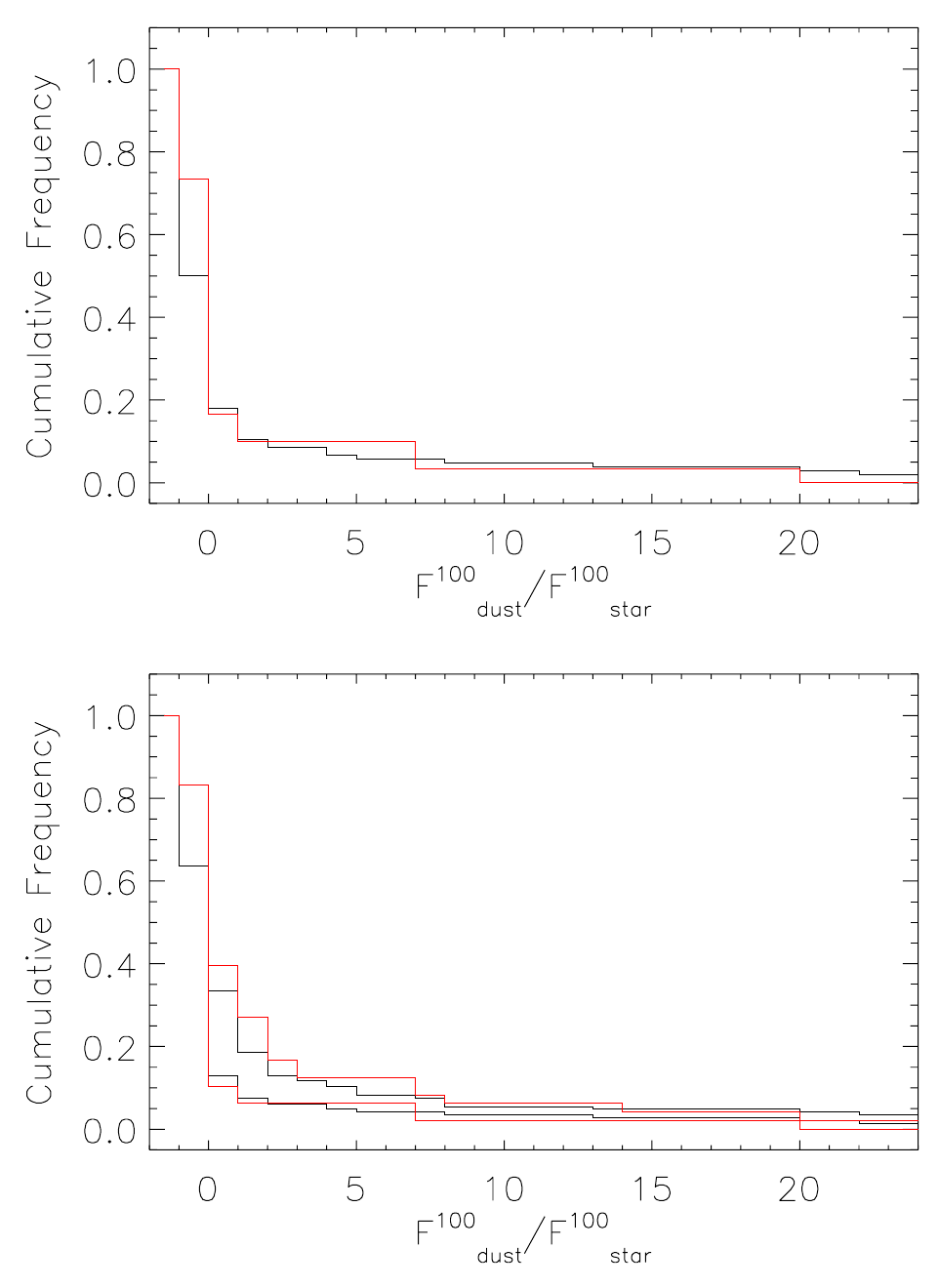}
\caption{Cumulative frequency of the dust flux ratio at 100 $\mu$m. {\bf Top}: only for the stars with significant detected emission (i.e., $F_{100}/\sigma_{100} >$ 3 -- this panel is biased to large excesses because for stars with faint photospheres, they can be included only if the have large dust flux ratios). {\bf Bottom}:  for all the stars assuming an optimistic case, where the adopted flux ratio for the targets without significant detected emission is its corresponding upper limit, and a pessimistic case, where the adopted flux ratio is 0. Black is for the stars with ages $>$ 1Gyr (Set 1o) and red is for stars with ages $<$ 1Gyr (Set 1y).}
\label{fr_100Myr_cum_det_opt_pes_age}
\end{figure}

\begin{figure}
\centering
\includegraphics[width=9cm]{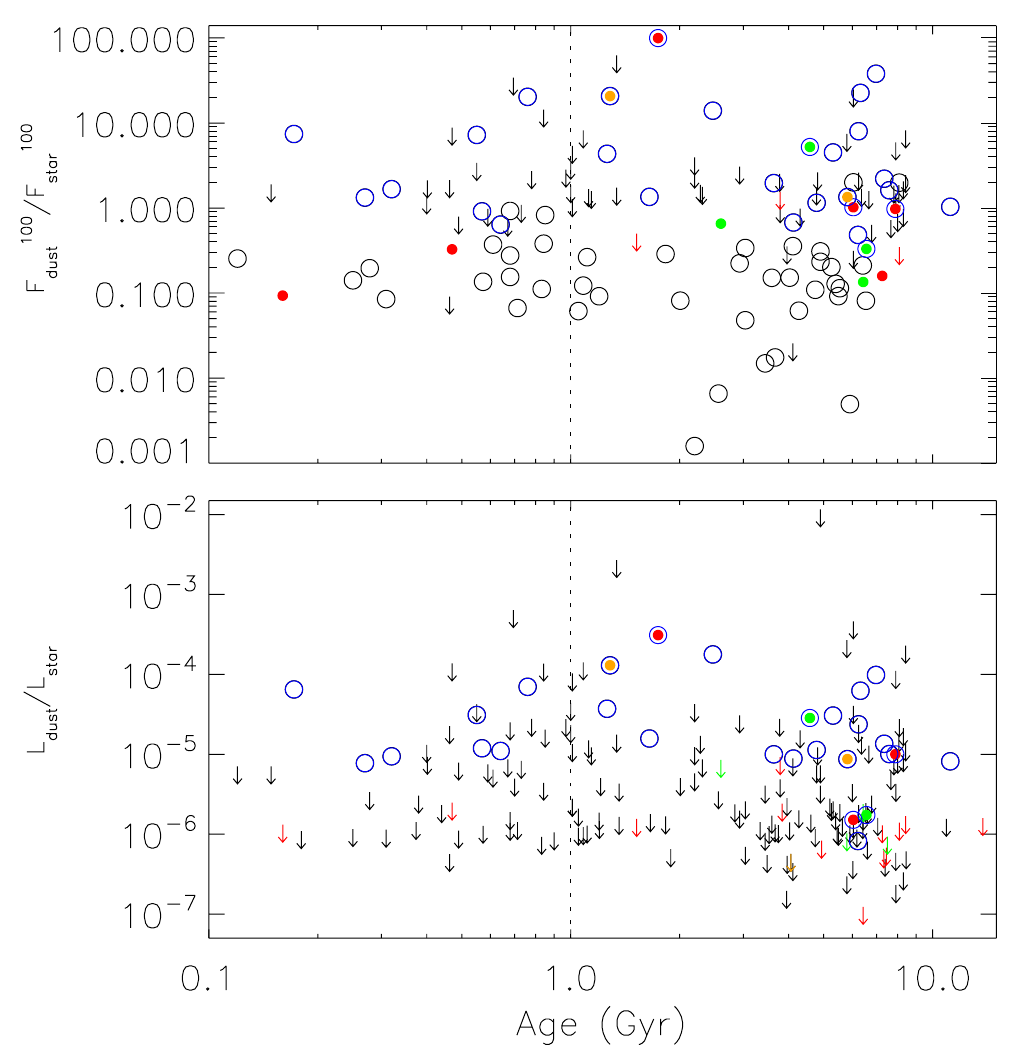}
\caption{{\bf Top}: Dust flux ratio at 100 $\mu$m as a function of stellar age. The circles correspond to detections (i.e., $F_{100}/\sigma_{100} >$ 3), while the down-facing arrows correspond to upper limits (i.e., $F_{100}/\sigma(F_{100}) <$ 3). 
Black is for the stars without known planets (Set 2), red is for the high-mass planet sample (Set 3), and green is for the low-mass planet sample  (Set 4). Unconfirmed planetary systems appear in orange. The larger open blue circles indicate which of those stars harbor excess emission at 100 $\mu$m (Set 5).
{\bf Bottom}: Same as above but for the fractional luminosity, assuming a blackbody emission from the excess. The circles correspond to dust detections (i.e., stars with SNR$_{\rm dust} > 3$, where SNR$_{\rm dust} = {F_{\rm obs}^{\rm100}-F_{\rm star}^{\rm100} \over \sqrt{{\sigma_{\rm obs}^{\rm100}}^2+{\sigma_{\rm star}^{\rm100}}^2}}$), while the down-facing arrows correspond to upper limits (i.e., SNR$_{\rm dust} < 3$).}
\label{fr_lr_age}
\end{figure}

Figure \ref{fr_100Myr_cum_det_opt_pes_age} shows the cumulative distribution of the dust flux ratio, whereas Figure \ref{fr_lr_age} shows its dependency with stellar age. To assess quantitatively whether the data show a decay with time, we carry out survival analysis. This is favored over the Kolmogorov-Smirnov (K-S) test because the latter does not deal with upper limits, and a significant number of the targeted stars have $F_{100}/\sigma_{100} <$ 3 (see Table \ref{tab:obs1} and down-facing arrows in Figure \ref{fr_lr_age}). Using ASURV 1.2 (Lavalley et al. \citeyear{1992ASPC...25..245L}), which implements the survival analysis methods of Feigelson \& Nelson (\citeyear{1985ApJ...293..192F}), we carried out the univariate, nonparametric two-sample Gehan, logrank, and Peto-Prentice tests to compute the probability that Sets 1y and 1o have been drawn from the same parent distribution with respect to the dust flux ratio. The results are listed in Table \ref{tab:stat_result}, line 5. The logrank test is more sensitive to differences at low values of the variable (i.e., near the upper limits), whereas the Gehan test is more sensitive to differences at the high end (i.e., at the detections; Feigelson \& Nelson \citeyear{1985ApJ...293..192F}). The Peto-Prentice test is preferred when the upper limits dominate and the sizes of the samples to be compared differ. The probabilities are not low enough to claim definitively that the two sets have been drawn from different distributions in terms of the dust flux ratio. However, given that they are in the 3--11\% range to assess the role of planet presence, we will take the conservative approach of limiting the comparison of disk frequencies and dust flux ratios to stars with ages $>$ 1 Gyr (i.e., within Set 1o). 

\subsection {Dependence on planet presence}
\label{sec:planets}

\subsubsection{High-mass planets}
\label{sec:highmasspl}

To assess the effect of high-mass planets on the presence of debris disks, we compare the disk frequencies in Set 3o (3/14 = 0.21) and Set 2o (16/126 = 0.13), limiting, for the reasons explained above, the comparison to the stars older than 1Gyr. Using a binomial distribution, we find that detecting three or more disks in Set 3o when the expected detection rate is 0.13 (taking Set 2o as reference, i.e. the expected number of disk detections is 0.13$\cdot$14) is a 27\% event; the probability drops to 9\% if including the unconfirmed planetary systems (Table \ref{tab:stat_result}, lines 9 and 10). Based on these numbers, there is no evidence that debris disks are more common around stars harboring high-mass planets compared to the average population, in agreement with previous studies based on {\it Spitzer} observations (Moro-Mart\'{\i}n\ et al. \citeyear{2007ApJ...658.1312M}; Bryden et al. \citeyear{2009ApJ...705.1226B}; K\'osp\'al et al. \citeyear{2009ApJ...700L..73K}). 

Classifying the stars in both samples (Sets 2o and 3o) into stars with and without disks and using the Fisher exact test, we find that there is a 41\% probability to find the observed arrangement of the data if the null hypothesis were true, where the null hypothesis in this case is that the stars with at least one giant planet (Set 3o) and the stars without known-planet planets (Set 2o) are equally likely to harbor disks. This probability is 11\% if including the unconfirmed planetary systems (Table \ref{tab:stat_result}, lines 11 and 12).  The Fisher exact test, therefore, does not identify any correlation between debris disk frequency and high-mass planet presence. To test how different the disk frequencies would have to be for a correlation to be identified by the Fisher exact test, we carry out the test using Set 2o and a hypothetical Set 3o, varying in the latter the number of stars with and without disks: we find that the disk frequency for Set 3o would have to be about 2.8 times higher than in Set 2o. The identification of smaller differences in disk frequencies by the Fisher exact test is limited by low-number statistics. 

\begin{figure}
\centering
\includegraphics[width=7cm]{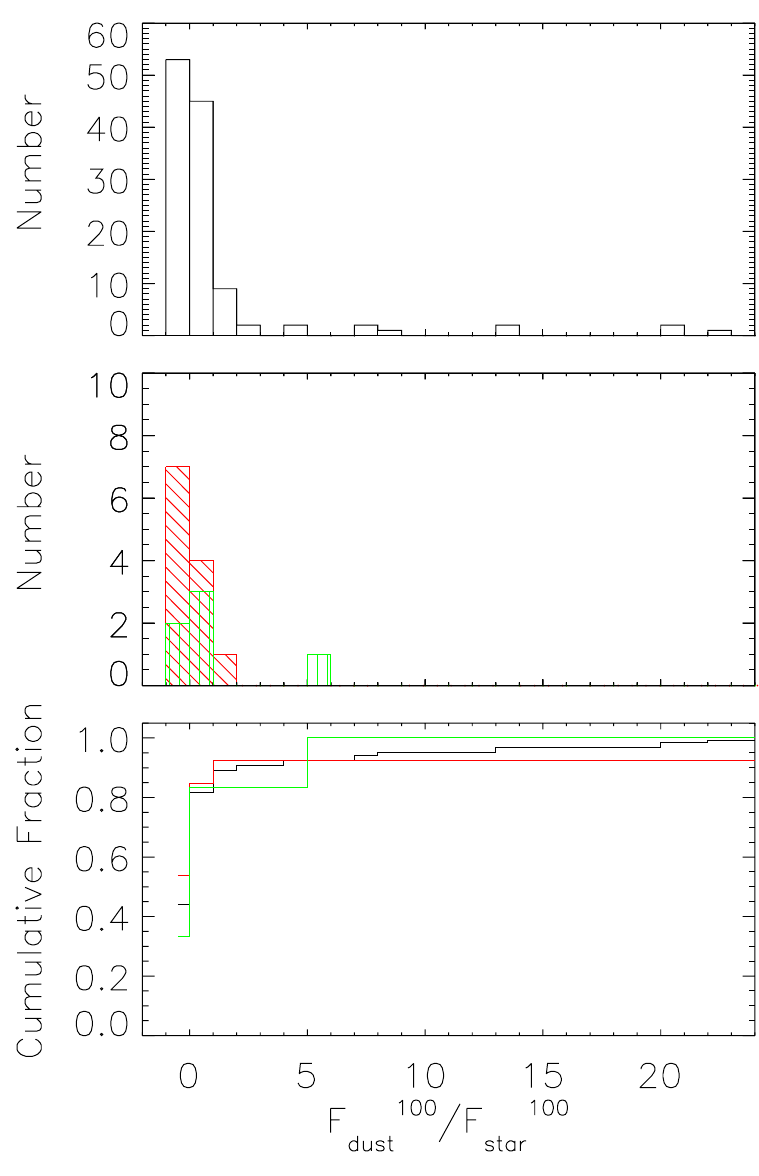}
\caption{Distribution of the excess flux ratio at 100 $\mu$m for stars with significant detected emission (i.e., $F_{100}/\sigma_{100} >$ 3). {\bf Top}: The open black histogram corresponds to the stars without known planets (Set 2). {\bf Middle}: The red filled histogram (with hatching from the top-left to the bottom right) corresponds to the high-mass planet sample (Set 3), while the green filled histogram (with vertical hatching) to the low-mass planet sample (Set 4). 
{\bf Bottom}: Cumulative fraction (same color code as above). There are two stars outside the plotted range, one in Set 3a with $F_{\rm dust}^{\rm100}/F_{\rm star}^{\rm100}$ = 99.8 and another in Set 2 with $F_{\rm dust}^{\rm100}/F_{\rm star}^{\rm100}$ = 38.0. 
}
\label{fr_fr_det_100Myr_cum}
\end{figure}

\begin{figure}
\centering
\includegraphics[width=9cm]{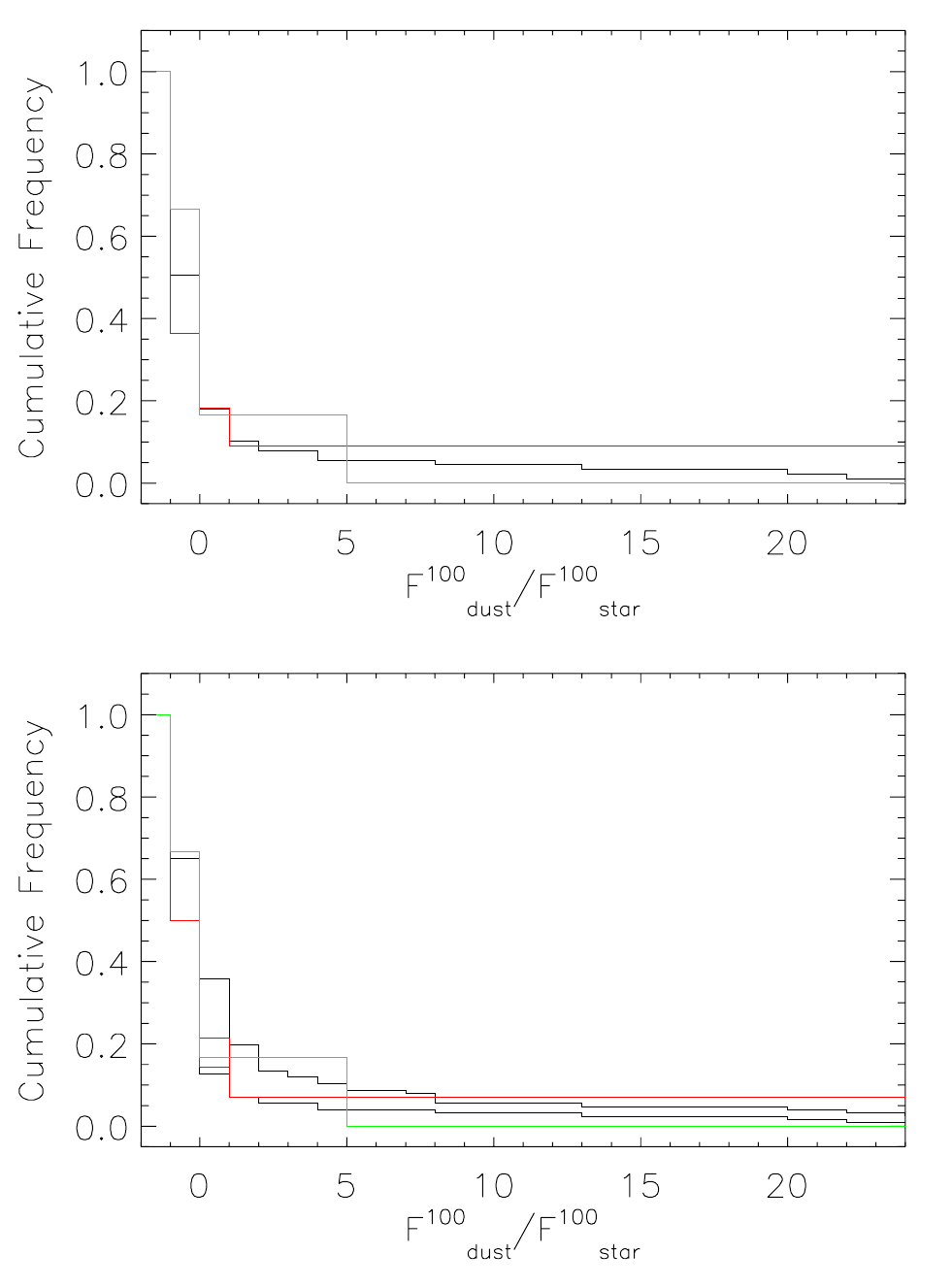}
\caption{Cumulative frequency of the dust flux ratio at 100 $\mu$m. {\bf Top}: only for the stars with significant detected emission (i.e., $F_{100}/\sigma_{100} >$ 3). {\bf Bottom}:  for all the stars assuming an optimistic case, where the adopted flux ratio for the targets without significant detected emission is its corresponding upper limit, and a pessimistic case, where the adopted flux ratio is 0. Black is for the stars without known planets with ages $>$ 1Gyr (Set 2o), red for stars harboring high-mass planets (Set 3o) and green for those harboring low-mass planets (Set 4o). The unconfirmed planetary systems are included under Set 2 (no-planet sample).}
\label{fr_100Myr_cum_det_opt_pes}
\end{figure}

Using survival analysis, we address whether the dust flux ratio, $F_{\rm dust}^{\rm100}/F_{\rm star}$, is affected by the presence of high-mass planets. Figures \ref{fr_fr_det_100Myr_cum} and \ref{fr_100Myr_cum_det_opt_pes} show the distribution of the dust flux ratio. The results from survival analysis (Table \ref{tab:stat_result} -- lines 15 and 16) indicate that there is a high probability that the high-mass planet sample (Set 3o) and the no-planet sample (Set 2o) have been drawn from the same population in terms of the dust flux ratio at 100 $\mu$m (and the result holds if we include the unconfirmed planetary systems). The data do not show evidence that the disks around high-mass planet-bearing stars harbor more dust than those without known planets but with similar stellar characteristics.  

\subsubsection {Low-mass planets}
\label{sec:lowmasspl}

We now repeat the exercise above for the low-mass planet sample, comparing the disk frequency in Set 4o (2/6 = 0.33) to that in Set 2o (16/126 = 0.13). Using a binomial distribution, we find that detecting two or more disks in Set 4o when the expected number of disk detections is 0.13$\cdot$6 (taking Set 2o as reference) is a 18\% probability event; the probability drops to 5\% if including the unconfirmed planetary systems (Table \ref{tab:stat_result}, lines 17 and 18). Based on these numbers there is no firm evidence that debris disks are more common around stars harboring low-mass planets compared to the average population. This test, however, does not take into account the uncertainty in the expected rate of the reference sample.

The Fisher exact test gives in a 19\% probability to find the observed arrangement of the data if the null hypothesis were true, where the null hypothesis is that the stars with low-mass planets only (Set 4o) and the stars without planets  (Set 2o) are equally likely to harbor disks.   The probability drops to 7\% when including the unconfirmed planetary systems (Table \ref{tab:stat_result}, lines 19 and 17). We find that the disk frequency for Set 4o would have to be about four times higher than in Set 2o in order for the Fisher exact test to identify a correlation in our small subsample Set 4o. The identification of smaller differences in disk frequencies is limited by low-number statistics.


The results from survival analysis (Table \ref{tab:stat_result} -- lines 23 and 24) indicate that the probability that the low-mass planet sample (Set 4o) and the no-planet sample (Set 2o) have been drawn from the same population in terms of the dust flux ratio at 100 $\mu$m is not low enough to claim a correlation (even when including the unconfirmed planetary systems). However, in this case,  survival analysis might be unreliable because of the small sample size (N$\lesssim$10) of the low-mass planet sample. 

In section \ref{sec:biases_sptype} below we discuss that there are hints that the debris disk frequency around F-type stars might be higher than around G- and K-type, although this trend is not found to be statistically significant.  However, given that none of the F-type stars in our sample harbor planets (see Figure \ref{SpT_histo}, because it is not possible to search to such low masses around them), to be conservative we now compare the low-mass planet sample to a control sample that does not include F-type stars. We find that the binomial-derived probability that the disk frequencies of the low-mass planet sample and the no-planet sample (excluding the F's) are similar is 9\% (compared to 14\% when including the F's). The Fisher exact probability gives 12\% (compared to 19\% when including the F's). Therefore, our conclusion that there is no evidence of correlation does not change when excluding F-type stars.

In summary, our study does not show evidence of a correlation, but our conclusion is limited by the small sample size. 

\subsubsection {Planetary system multiplicity}
\label{sec:multi}
 
Comparing Set 6o (single-planet sample, with a disk frequency of 0.3) and Set 7o (multiple-planet sample, with a disk frequency of 2/10=0.20) and using a binomial distribution, we find that detecting two or more disks in Set 7o when the expected detection rate is 0.30 (taking Set 6o as reference, i.e. the expected number of disk detections is 0.30$\cdot$10) is an 85\% probability event (changing only slightly when including the unconfirmed planetary systems -- Table \ref{tab:stat_result}, lines 25 and 26). The data do not show any evidence that debris disks are more or less common around stars harboring multiple-planet systems compared to single-planet systems.  The same conclusion results from the Fisher exact test (Table \ref{tab:stat_result}, lines 27 and 28). Regarding the dust flux ratio,  survival survival analysis results (Table \ref{tab:stat_result}, lines 29--34) indicate that the multiple-planet,  single-planet and no-planet samples could have been drawn from the same population in terms of the dust flux ratio at 100 $\mu$m (and the result holds if we include the unconfirmed planetary systems). The data, again, do not show evidence of any correlation between planet multiplicity and the strength of the debris disk emission. 
 
\subsubsection {Effect on the characteristic dust temperature}
\label{sec:results_tdust}

\begin{figure}
\centering
\includegraphics[width=8cm]{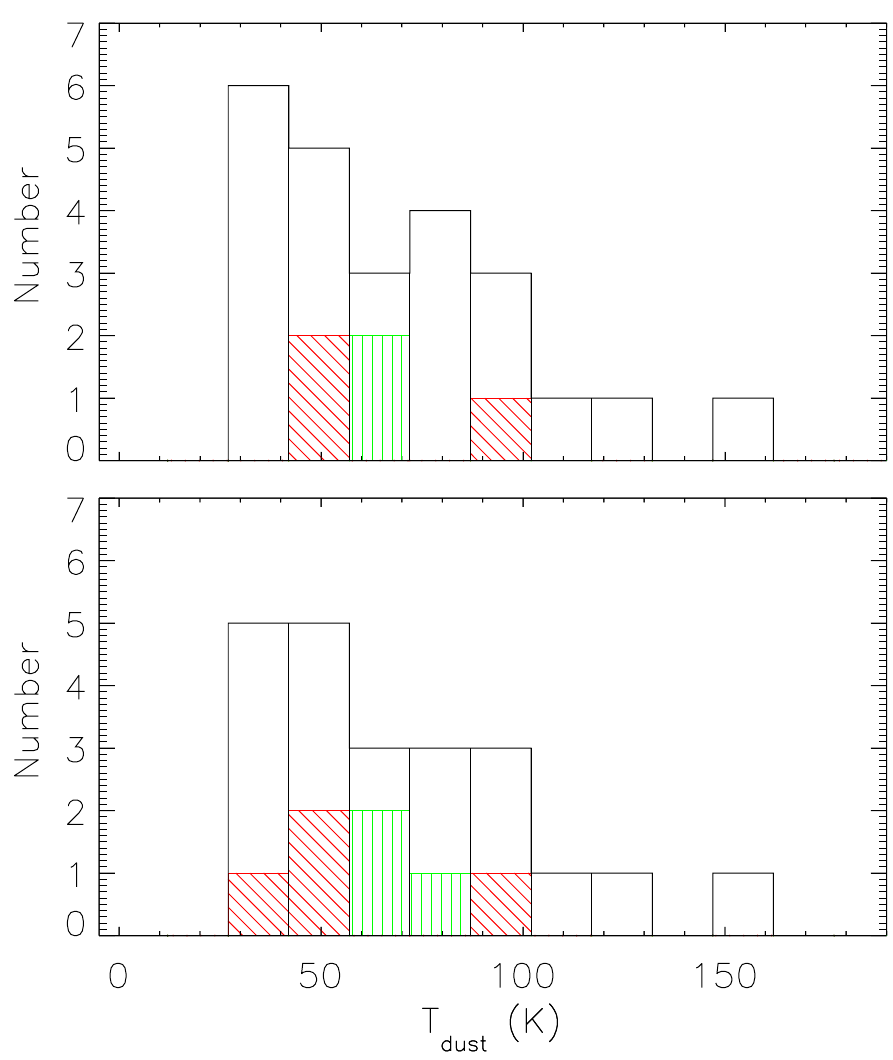}
\caption{Distribution of the estimated black-body dust temperature for the stars with debris disk detections at 100 $\mu$m  (i.e., SNR$_{\rm dust} > 3$). The open black histogram corresponds to stars without known planets (Set 2t), while the line-filled colored histogram corresponds to stars harboring high-mass planets (Set 3t; in red, with hatching from the top-left to the bottom right) and stars harboring low-mass planets (Set 4t; in green, with vertical hatching). The {\bf top} panel excludes unconfirmed planetary systems $\epsilon$ Eri and $\tau$ Cet, while the {\bf bottom} panel includes both planetary systems.}
\label{tdust_histo}
\end{figure}

We now assess whether there is any evidence that the debris disks around planet-bearing stars might be different from those around an average population of stars in terms of the characteristic dust temperature. Sets labeled with a "t" include only the stars with estimated dust temperatures (listed in Table \ref{tab:diskprop}). The calculation of the gray-body dust temperatures is described in Kennedy et al. (\citeyear{2012MNRAS.426.2115K}) based on observations with a wide wavelength coverage. Figure \ref{tdust_histo} shows the distribution of the characteristic dust temperature in the no-planet sample (Set 2t) and the planet samples (Sets 3t and 4t). The K-S test yields two values, $D$, a measure of the largest difference between the two cumulative distributions under consideration, and the probability of finding a $D$-value greater than the observed value; the latter is an estimate of the significance level of the observed value of $D$ as a disproof of the null hypothesis that the distributions come from the same parent population; that is, a small probability implies that the distributions could be significantly different. The result from the K-S test is shown in Table \ref{tab:stat_result} (lines 35 and 36), showing a very high probability. The calculation of the probability is good if N$_1$N$_2$/(N$_1$+N$_2$) $\ge$ 4, where N$_1$ and N$_2$ are the number of stars in each set. However, if one wants to be conservative, it might be compromised when N $<$ 20, as it is the case here. Based on the limited information we have so far, there is no evidence that the characteristic temperature of the debris disks around planet-bearing stars differs from the rest.  

\section {Correlations with stellar metallicity}
\label{sec:results_met}

\begin{figure}
\centering
\includegraphics[width=8cm]{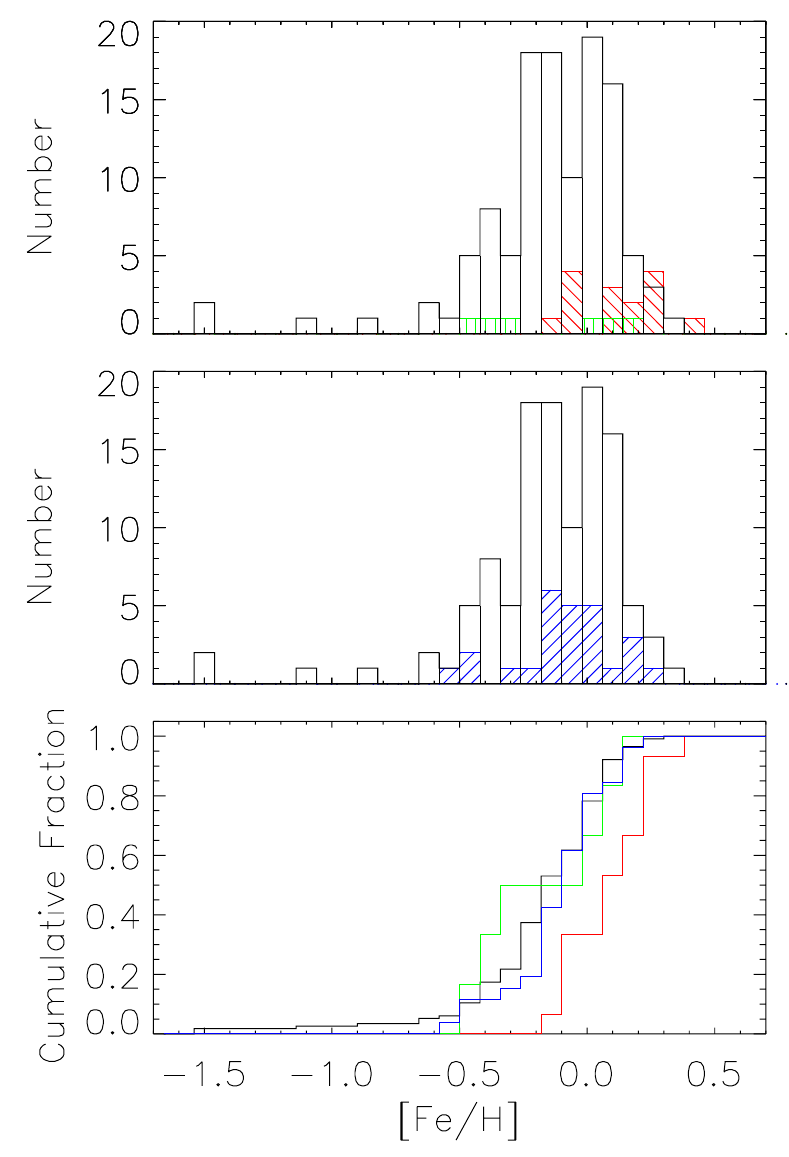}
\caption{Distribution of stellar metallicities (logarithmic scale, with [Fe/H] = 0.0 for solar metallicity). The open black histograms correspond to stars without known planets and with known metallicities (Set 2m). {\bf Top}: The line-filled colored histograms  correspond to stars harboring high-mass planets (Set 3m; in red, with hatching from the top-left to the bottom right), and low-mass planets  (Set 4m; in green, with vertical hatching).  {\bf Middle}: subset harboring  excess emission at 100 $\mu$m (Set 5m; in blue, with hatching from the bottom-left to the top-right). The stars with unconfirmed planetary systems, $\epsilon$ Eri and $\tau$ Cet, are included in Set 2m (no-planet sample). {\bf Bottom}: Cumulative distributions of stellar metallicities (same color code as above). 
}
\label{met_histo_cum}
\end{figure}

Figure \ref{met_histo_cum} shows the distribution of stellar metallicity.  To assess the correlation with metallicity, we create Sets 1m--5m, constituted by stars in Sets 1--5 with known metallicities\footnote{Regarding possible sources of biases due to stellar age and distance, Maldonado et al. (\citeyear{2012A&A...541A..40M}) argued that because the stars are at close distances from the Sun (in our case within 20 pc), it is unlikely that they have suffered different enrichment histories.} from Maldonado et al. (\citeyear{2012A&A...541A..40M}) and Eiroa et al. (\citeyear{2013A&A...555A..11E}). These sets are further divided into stars with high metallicities (Sets 1h--5h) and those with low metallicities (Sets 1l--5l), using the midpoint of the metallicity distribution, [Fe/H] = -0.12, as the dividing value. Table \ref{tab:met} lists how many stars are in each subset. 

\subsection{Debris disk presence}

We now compare the debris disk frequencies in Set 1h (17/75 =  0.23) and Set 1l (9/61 = 0.15). Using a binomial distribution, finding 17 or more disk detections in Set 1h, when the expected detection rate is 0.15 (taking Set 1l as reference,  i.e. the expected number of disk detections is 0.15$\cdot$75) is a 4\% probability event (Table \ref{tab:stat_result} -- line 37), indicating that the disk frequencies in the high- and low-metallicity samples might differ. This result, however, does not take into account the uncertainty in the expected rate of the reference sample (in this case Set 1l). From the Fisher exact test, we find that there is a 28\% probability to find the observed arrangement of the data if the null hypothesis were true, where the null hypothesis in this case is that the stars without disks (Set 1m-Set 5m) and the stars with disks (Set 5m) are equality likely to have metallicities $>$ -0.12 (Table \ref{tab:stat_result} -- line 38).  From the K-S test, the probability that the no-planet sample (Set 2m) and the debris disk sample (Set 5m) could have been drawn from the same distribution in terms of stellar metallicity is 33\% (39\% when including unconfirmed planetary systems; Table \ref{tab:stat_result} -- lines 47--48). 

\begin{figure}
\centering
\includegraphics[width=8cm]{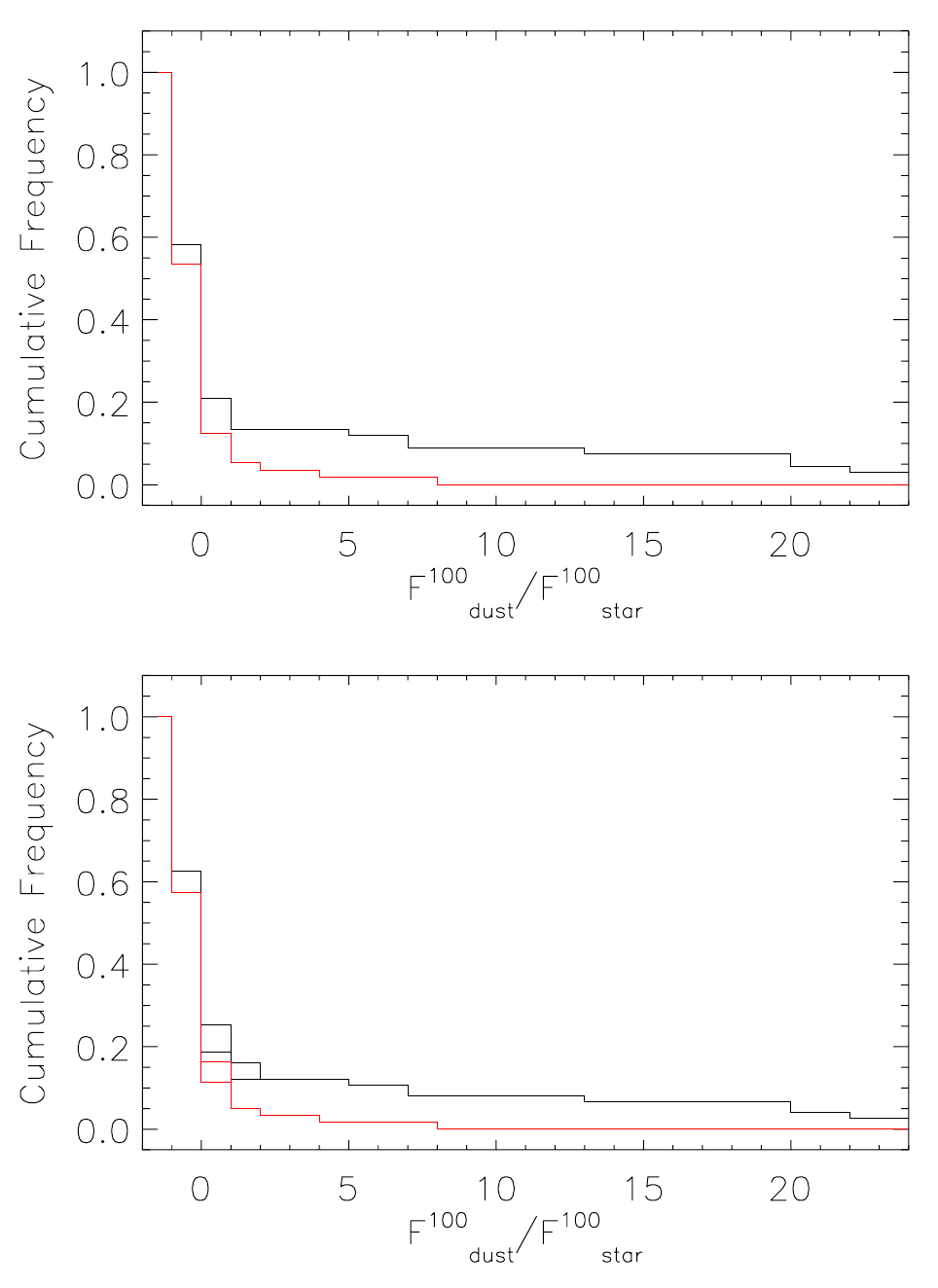}
\caption{Cumulative frequency of the dust flux ratio at 100 $\mu$m. {\bf Top}: only for the stars with significant detected emission (i.e., $F_{100}/\sigma_{100} >$ 3). {\bf Bottom}:  for all the stars assuming an optimistic case, where the adopted flux ratio for the targets without significant detected emission is its corresponding upper limit, and a pessimistic case, where the adopted flux ratio is 0. Black is for the stars with metallicities larger than the average ([Fe/H] $>$ -0.12; Set 1h) and red is for the stars with lower metallicities ([Fe/H] $\leqslant$ -0.12; Set 1l), independently of planet presence.}
\label{fr_100Myr_cum_det_opt_pes_met}
\end{figure}

\begin{figure}
\centering
\includegraphics[width=8cm]{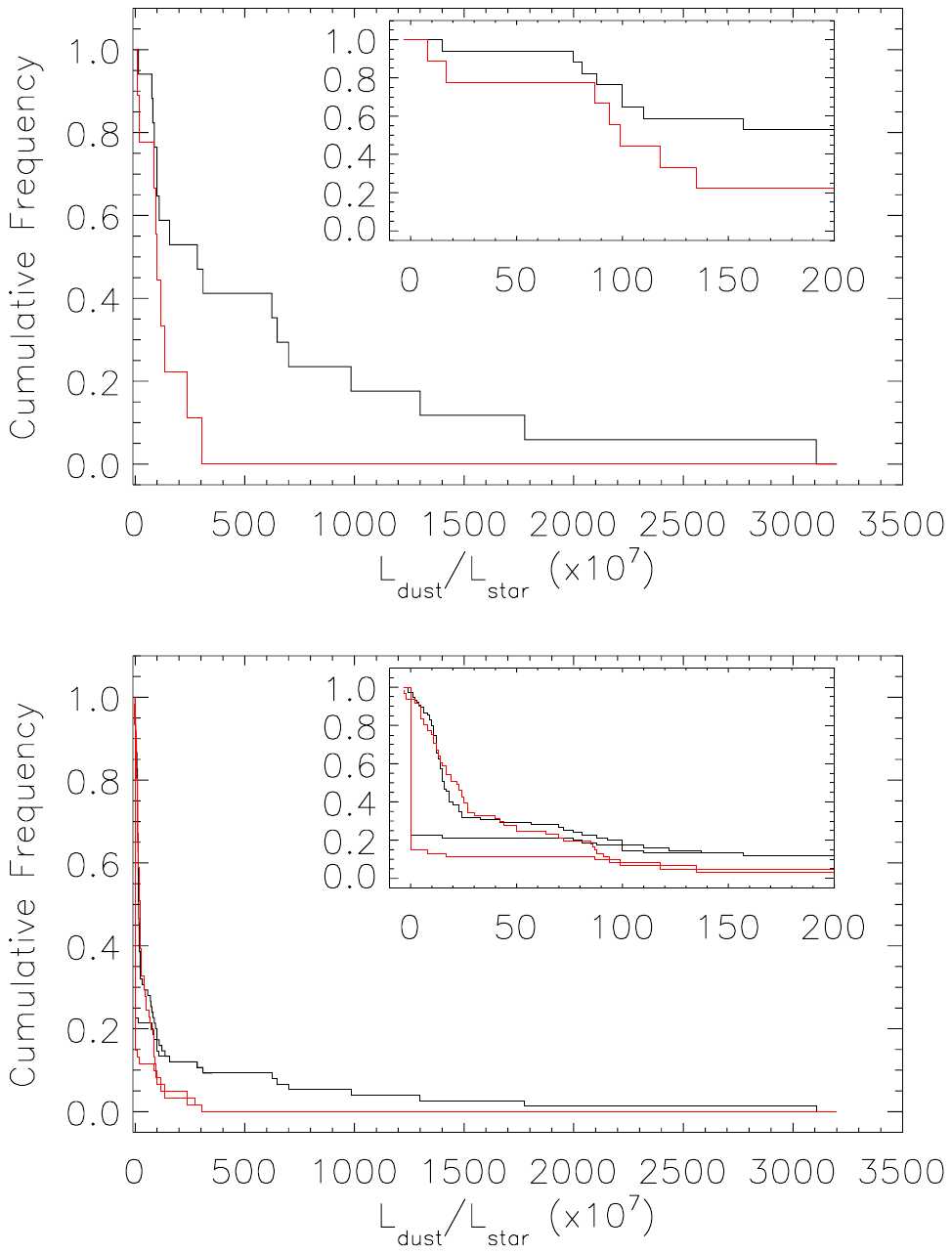}
\caption{Cumulative frequency of the dust fractional luminosity. {\bf Top}: only for the stars with excess detections (i.e., stars with SNR$_{\rm dust} > 3$).  {\bf Bottom}:  for all the stars assuming an optimistic case, where the adopted fractional luminosity for the targets without excess detections is its corresponding upper limit, and a pessimistic case, where the adopted fractional luminosity is 0. Black is for the stars with metallicities larger than the average ([Fe/H] $>$ -0.12; Set 1h) and red is for the stars with lower metallicities ([Fe/H] $\leqslant$ -0.12; Set 1l), independently of planet presence.}
\label{lr_100Myr_cum_det_opt_pes_met}
\end{figure}

Regarding the strength of the excess emission, we use survival analysis to check if the low-metallicity and high-metallicity samples could have been drawn from the same population in terms of the dust flux ratio. Figures \ref{fr_100Myr_cum_det_opt_pes_met} and Figure \ref{lr_100Myr_cum_det_opt_pes_met} show the cumulative frequencies of the dust flux ratio and the fractional luminosity of Sets 1h and 1l, showing that there is a dearth of debris disks with high dust flux ratios and high fractional luminosities around low-metallicity stars. However, the probabilities listed in Table \ref{tab:stat_result} (line 49) indicate that this trend is not statistically significant. We cannot rule out the hypothesis that the high-metallicity and low-metallicity samples have been drawn from the same distribution in terms of the dust flux ratio. We conclude that the Fisher exact test and survival analysis do not allow us to identify any correlation between high stellar metallicity and debris disks. 

\subsection{Planet presence}

Comparing the planet and no-planet samples in terms of stellar metallicity with the Fisher exact test (Table \ref{tab:stat_result} -- lines 39--42), we find that in the case of giant planets, there is a 0.2\% probability to find the observed arrangement if the stars without giant planets (Set 1m-Set 3m) and the stars with giant planets (Set 3m) were equally likely to have metallicities $>$ -0.12, whereas for low-mass planets (Set 1m-Set 4 vs. Set 4) this probability is almost 100\% (the result holds when including the unconfirmed planetary systems). From the K-S test, the probability that the no-planet sample and the high-mass planet sample could have been drawn from the same distribution in terms of stellar metallicity is 0.2\%, whereas the probability that the no-planet sample has been drawn from the same distribution as the low-mass planet sample and the debris disk sample is much larger (49\%; Table \ref{tab:stat_result} -- lines 43--46). 

\section {Possible biases introduced by the sample selection}
\label{sec:biases}

\subsection{Presence of undetected planets}
\label{sec:biases_pl}

We now describe the potential biases that the sample selection could introduce in the statistical analysis described above. First, we assess whether the presence of unidentified planetary systems could affect our results. If we were to have many stars with high-mass planets in the control sample, Set 2, one could argue that a high-mass planet-debris disk correlation could have been present but hidden by all the  ``planet contaminants". However, because the high-mass planet frequency is small, this seems unlikely. Due to the higher frequency of low-mass planets (Mayor et al. \citeyear{2009A&A...493..639M}, \citeyear{2011arXiv1109.2497M}; Batalha \citeyear{2014PNAS..11112647B} and references therein; Marcy et al. \citeyear{2014PNAS..11112655M} and references therein), we probably have many stars with low-mass planets in the control sample which have not been identified. This means that a low-mass planet-debris disk correlation may still be hidden in the data. We could avoid these biases by comparing the planet sets to a subset of stars in Set 2 for which the presence of planets within a given period and mass has been ruled out by the radial velocity surveys. However, because nondetections are generally not made public by the planet search teams, the information to construct this no-planet stellar sample is not available. 

\subsection{Distribution of spectral types}
\label{sec:biases_sptype}

By considering FGK stars to assess the planet-debris disk correlation, we are implicitly assuming that the disk frequency and the planet frequency do not differ significantly among these spectral types. 

\begin{figure}
\begin{center}
\includegraphics[width=5cm]{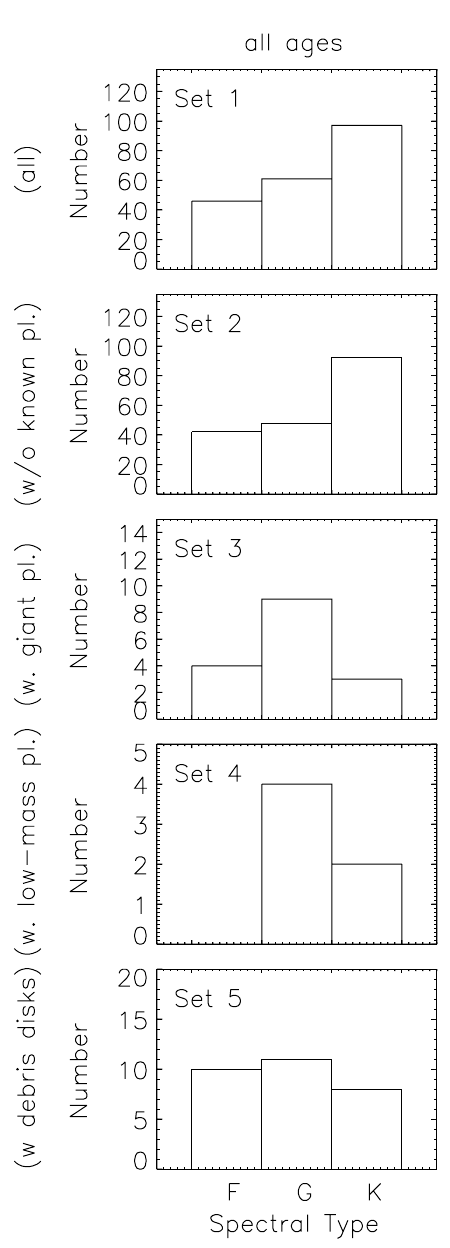}
\end{center}
\caption{Distribution of spectral types for the different sets.}
\label{SpT_histo}
\end{figure}

Table \ref{tab:sptype} and Figure \ref{SpT_histo} show the distribution of spectral types in the samples under consideration.  Let us limit the comparison to stars older than 1 Gyr (to avoid biases due to disk evolution), i.e. to the stars in Set 1o (Table \ref{tab:sptype} -- line 6). For the F-stars, the disk fraction is 0.24 (8/33 disk detections): using a binomial distribution, finding eight or more disk detections, when the expected detection rate is 0.14 (taking the disk frequency of the G-stars as reference, i.e. when the expected number of disk detections is 0.14$\cdot$33) is an 8\% probability event. While for the K-stars, with a  disk fraction of 0.09 (6/64 disk detections), using a binomial distribution, finding six or more disk detections,  when the expected detection rate is 0.14 is a 90\% probability event. If we were to take the disk frequency of K-type as reference, for the F-stars, finding eight or more disk detections,  when the expected detection rate is 0.09 (expected number of disk detections of 0.09$\cdot$33) would be a 0.8\% event (Table \ref{tab:stat_result} -- lines 50--52). The latter seems to indicate there is a significant difference in disk frequencies between K-type and F-type stars. 

Eiroa et al. (\citeyear{2013A&A...555A..11E}) found that the frequency of disks in the {\it DUNES} survey does not change significantly among FGK stars. The increased disk frequency for F-type stars found in our sample might have been biased  to some degree by the shallower integration time of some of the {\it DEBRIS} targets, although the different T$_{\rm eff}$ distribution for the stars in the {\it DEBRIS} and {\it DUNES} surveys may also play a role (the former covering all FGK stars, whereas the latter covers mid-F to mid-K\footnote{The spectral type dependence of the debris disk frequency within the {\it DEBRIS} sample will be studied in more detail by Sibthorpe et al. (in preparation).}. Using a larger sample of {\it Spitzer} and {\it Herschel} observations, Sierchio et al. (\citeyear{2014ApJ...785...33S}) found no significant dependence with spectral type in the F4-K4 range.

The test above does not consider the uncertainty in the expected rate of the reference sample. Classifying the stars into those with and without debris disks and applying the Fisher exact test, we find that in this case the probability is not low enough to disprove the null hypothesis that the F-stars are equally likely to harbor disks as are the G+K stars (Table \ref{tab:stat_result} -- lines 53 and 54). 

Regarding planet frequency, Doppler surveys indicate there is a correlation between high-mass planet frequency and spectral type that follows roughly a linear increase with stellar mass (Johnson et al. \citeyear{2010PASP..122..905J}). From a compilation of Doppler surveys, Gaidos et al. (\citeyear{2013ApJ...771...18G}) suggest f(\%) = -1.11 + 5.33 M$_{star}$/M$_\odot$, for planets $>$ 8 R$_{\oplus}$ (masses $>$ 95 M$_{\oplus}$ -- see their Figure 8). For low-mass planets in the 0.8--6R$_{\oplus}$ range, Kepler data indicate that among the FGK stars the planet frequency does not depend significantly on the spectral type (Fressin et al. \citeyear{2013ApJ...766...81F}). Table \ref{tab:sptype} and Figure \ref{SpT_histo} indicate that neither the high-mass nor low-mass planet frequencies within our sample reflect the above trends, with a higher incidence around G-type stars mostly likely because fewer F and Ks were searched for planets. This might skew slightly the disk incidence rate comparison for high-mass planets. Again, because nondetections are generally not made public, there is no way to circumvent this issue. 

In Section \ref{sec:summary_met} we discuss how the conclusions change when excluding F-type stars from our analysis. 

\section{Fractional luminosities and comparison to the Solar system's debris disk}
\label{sec:frac_lum}

\begin{figure}
\centering
\includegraphics[width=8.5cm]{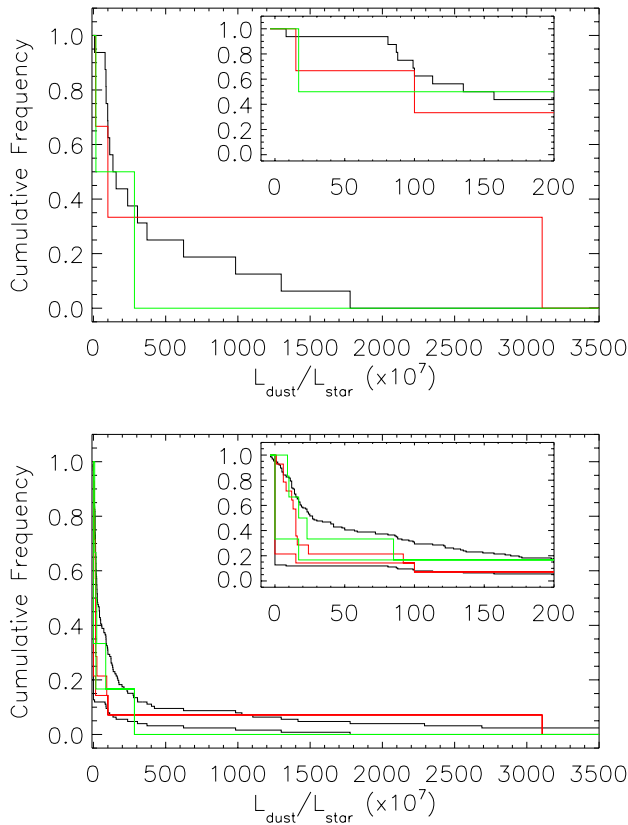}
\caption{Cumulative frequency of the dust fractional luminosity. {\bf Top}: only for the stars with excess detections (i.e., stars with SNR$_{\rm dust} > 3$). {\bf Bottom}:  for all the stars assuming an optimistic case, where the adopted fractional luminosity for the targets without excess detections is its corresponding upper limit, and a pessimistic case, where the adopted fractional luminosity is 0. Black is for the stars without known planets with ages $>$ 1 Gyr (Set 2o), red for stars harboring high-mass planets (Set 3o) and green for those harboring low-mass planets (Set 4o). The unconfirmed planetary systems are included under Set 2 (no-planet sample).}
\label{lr_100Myr_cum_det_opt_pes}
\end{figure}

Figure \ref{lr_100Myr_cum_det_opt_pes} shows the cumulative frequency of the dust fractional luminosity. This variable is commonly used to characterize debris disk emission because it allows comparison of disks observed at different wavelengths; it is not very model-dependent as long as the wavelength coverage is good (as is the case in our samples). For stars with dust excess detections (SNR$_{\rm dust} > 3$), the fractional luminosity is calculated following Kennedy et al. (\citeyear{2012MNRAS.426.2115K}; \citeyear{2012MNRAS.421.2264K}). For stars with dust excess nondetections (SNR$_{\rm dust} < 3$), the 3$\sigma$ upper limit to the fractional luminosity is calculated from ${L_{\rm dust} \over L_{star}} = \big({T_{\rm dust} \over T_{star}}\big)^4 \big({e^{x_{\rm dust}}-1 \over e^{x_{\rm star}}-1}\big){F_{\rm obs}^{\rm100}-F_{\rm star}^{\rm100} \over F_{\rm star}^{\rm100}}$ following equation (4) in Beichman et al. (\citeyear{2006ApJ...652.1674B}), and assuming the observed flux is $F_{\rm obs}^{\rm100}+3\sigma_{\rm obs}^{\rm100}$, for stars with $F_{\rm obs}^{\rm100}/\sigma_{\rm obs}^{\rm100} > 0$, and 3$\sigma_{\rm obs}^{\rm100}$, for stars with $F_{\rm obs}^{\rm100}/\sigma_{\rm obs}^{\rm100} < 0$. In this expression, $x = {h\nu \over kT}$, where $\nu$ is the frequency corresponding to 100 $\mu$m, $T_{\rm star} = T_{\rm eff}$ is the effective stellar photospheric temperature, and $T_{\rm dust}$ is assumed to be 50 K (as in Eiroa et al. \citeyear{2013A&A...555A..11E}).

\begin{figure}
\centering
\includegraphics[width=8cm]{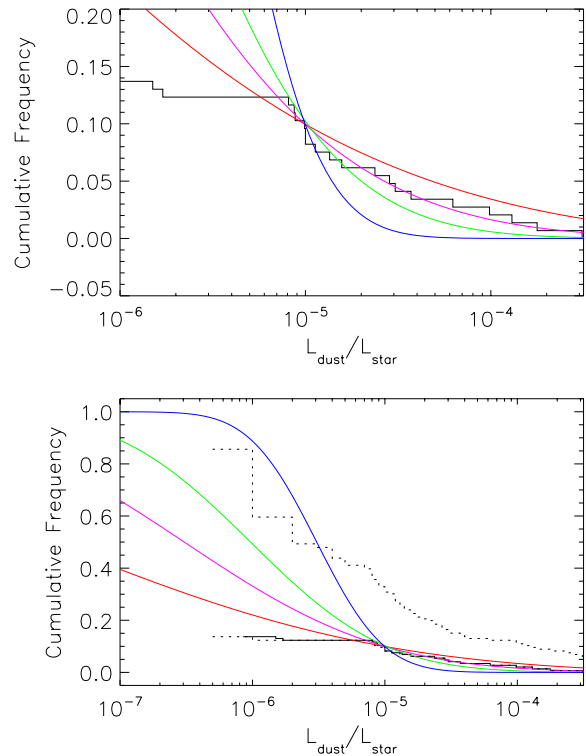}
\caption{Cumulative frequency of the fractional luminosity. The thick black histogram corresponds to the stars with ages $>$ 1 Gyr independently of planet presence (Set 1o). Because we are interested in the cumulative frequency of the stars for fractional luminosities greater than the minimum observed value, in calculating the cumulative distribution we adopt the pessimistic case, where the fractional luminosity for the stars without excess is 0.  The blue, green, magenta and red lines correspond to theoretical distributions that assume a Gaussian distribution of fractional luminosities in logarithmic scale, with average values of 10$\times$, 3$\times$, 1$\times$ and 0.1$\times$ that of the Solar system, respectively, and assuming for the Solar system a fractional luminosity of 10$^{-6.5}$.  We fixed the cumulative frequency of disks with L$_{\rm dust}$/L$_{*}$ $>$ 10$^{-5}$ at 10\% according to the observed result (in set 1o), implying 1-sigma widths for the theoretical distributions of 0.4, 0.8, 1.18 and 2.0, for the blue, green magenta and red lines respectively.  {\bf Top:} showing only the detected range; there are only three targets with fractional luminosities below 8$\cdot10^{-6}$, compromising the fit to the data in that low range because of low-number statistics. {\bf Bottom:} the dotted line that coincides with the solid line corresponds to the the pessimistic case, where the adopted fractional luminosities for the targets without excess detections are taken to be 0, while the second dotted line on the upper part of the panel corresponds to the optimistic case, for which the upper limits are adopted.}
\label{lr_set1o_SS}
\end{figure}

The fractional luminosity can help place the debris disk observations in this study in the context of the Solar system's debris disk. Following Bryden et al. (\citeyear{2006ApJ...636.1098B}), we compare the observed cumulative distribution of fractional luminosity to those expected from Gaussian distributions in logarithmic scale, with average values of 10$\times$, 3$\times$, 1$\times$, and 0.1$\times$ that of the Solar system's debris disk, assuming for the latter a fractional luminosity of 10$^{-6.5}$. To avoid biases due to disk evolution, we limit the comparison to stars older than 1 Gyr (Set 1o). The observed and Gaussian-derived cumulative distributions are shown in Figure \ref{lr_set1o_SS}. The bottom panel shows that the blue line exceeds the most optimistic case at low fractional luminosities. This means that we can reject the hypothesis that the median of the disk fractional luminosity is 10 times that of the Solar system's debris disk, in agreement with Bryden et al. (\citeyear{2006ApJ...636.1098B}). The best fit to the data is a Gaussian centered on the Solar system value (magenta line in the top panel). This result is discussed in Section \ref{sec:summary_ss}.

\section{Summary and Discussion}
\label{sec:summary}

We have carried out a statistical study of an unbiased subsample of the {\it Herschel} {\it DEBRIS} and {\it DUNES} surveys, consisting of 204 FGK stars located at distances $<$20 pc, with ages $>$ 100 Myr and with no binary companions at $<$100 AU. The main goal is to assess whether the frequency and properties of debris disks around a control sample of FGK stars are statistically different from those around stars with high-mass and low-mass planets. We find the following results. 

\subsection{Disk evolution} 
\label{sec:summary_evol}

The {\it Spitzer} surveys found that the upper envelope of the 70 $\mu$m debris disk emission shows a decline over the $\sim$ 100 Myr of a star's lifetime, indicating that there might be a population of rapidly evolving disks that disperse by 100 Myr. Our sample does not show clear evidence of disk evolution on the gigayear timescale. This is in agreement with the lack of disk evolution observed at 70 $\mu$m in {\it Spitzer} surveys for stars older than 1 Gyr\footnote{Compared to the 70 $\mu$m observations, the 100 $\mu$m emission might also trace dust located further out, where the collision times are longer; if this second population of dust exists, one would expect even less evolution at this longer wavelength.} (Trilling et al. \citeyear{2008ApJ...674.1086T}; Hillenbrand et al. \citeyear{2008ApJ...677..630H}; Carpenter et al. \citeyear{2009ApJS..181..197C}). In a recent study,  using both {\it Spitzer} and {\it Herschel} observations, and using a sample 2.5 times larger than ours, Sierchio et al. (\citeyear{2014ApJ...785...33S}) found that for disks with fractional luminosities smaller than 10$^{-5}$ there is a significant decrease in the debris disk frequency between 3 and 5 Gyr. To look for evidence of disk evolution in the 5 Gyr timescale that could bias our results, we have divided the sample into stars with ages 0.1--5 Gyr (labeled as Sets 1oy and 2oy) and stars older than 5 Gyr (Sets 1oo and 2oo). We then compare the disk frequencies and dust flux ratios in both subsamples (lines 6 and 8 in Table \ref{tab:stat_result}). The overall resulting probabilities  are not low enough to claim  that the two sets have been drawn from different distributions in terms of the dust flux ratio, nor that their disk incidence rates differ significantly. The Fisher eFisher exactxact test (line 7 in Table  \ref{tab:stat_result}) also indicates that both sets  are equally likely to harbor disks. We therefore do not find evidence in our restricted sample of disk evolution in the 5 Gyr timescale.

\subsection {High-mass planet presence}
\label{sec:summary_highmass}
Our sample do not show evidence that debris disks are more common around stars harboring high-mass planets compared to the average population. 
This is in agreement with the studies based on {\it Spitzer} observations that found no correlation between fractional luminosities, $L_{dust}/L_{star}$, and the presence of high-mass planets (Moro-Mart\'{\i}n\ et al. \citeyear{2007ApJ...658.1312M}; Bryden et al. \citeyear{2009ApJ...705.1226B}). Figure 8 in Maldonado et al. (\citeyear{2012A&A...541A..40M}) also shows this trend, where the stars with discs and planets seem to be well mixed with stars with only disks in terms of the fractional luminosity, but they did not carry out any statistical analysis. This issue will be revisited using a larger sample that combines {\it Herschel} {\it DEBRIS}, {\it DUNES}, and {\it SKARPS} observations (G. Bryden et al. 2015, in preparation).

Overall, the lack of observed correlation between high-mass planets and debris disks was understood within the context of the core accretion model for planet formation, where the conditions to form debris disks are more easily met than the conditions to form high-mass planets. This is in agreement with the metallicity studies that indicate that there is a correlation between high-mass planets and stellar metallicity, but no correlation between debris disks and stellar metallicity. Additionally, the presence of debris disks around stars with a very wide range of properties, from M-type (Kennedy et al. \citeyear{2007Ap&SS.311....9K}; Lestrade et al. \citeyear{2012A&A...548A..86L}) to the progenitors of white dwarfs (Jura \citeyear{2003ApJ...584L..91J}, \citeyear{2007ApJ...663.1285J}), implies that planetesimal formation is a robust process that can take place under a wide range of conditions. Therefore, based on formation conditions, if planetesimals can be common in systems with and without high-mass planets, there is no reason to expect a correlation between high-mass planets and debris disks (Moro-Mart\'{\i}n\ et al. \citeyear{2007ApJ...658.1312M}). 

Another factor contributing to the lack of a well-defined correlation with planet presence might be that the dynamical histories likely vary from system to system, and other stochastic effects need also to be taken into account, e.g.  those produced by dynamical instabilities of multiple-planet systems clearing the outer planetesimal belt (Raymond et al. \citeyear{2011A&A...530A..62R}, \citeyear{2012A&A...541A..11R}), the planetesimal belt itself triggering planet migration and instabilities (Tsiganis et al. \citeyear{2005Natur.435..459T}; Levison et al. \citeyear{2011AJ....142..152L}), or the stripping of planetesimals from disks during stellar flybys in the first 100 Myr, when systems are still in their dense birth cluster (Lestrade et al. \citeyear{2011A&A...532A.120L}). 

Another aspect that needs to be taken into account is that the planets detected by radial velocity surveys and the dust observed at 100 $\mu$m occupy well-separated regions of space, limiting the influence of the observed closer-in planets on the dust production rate of the outer planetesimal belt; there are long-range gravitational perturbations produced by secular perturbations from single planets on eccentric orbits (Mustill \& Wyatt \citeyear{2009MNRAS.399.1403M}) or multiplanet systems (Moro-Mart\'{\i}n\ et al. \citeyear{2007ApJ...668.1165M},  \citeyear{2010ApJ...717.1123M}) that allow close-in planets to excite outer planetesimal belts, but the timescale of the former may be longer than the age of the system, and the latter is limited to certain planet configurations. 

\subsection {Low-mass planet presence}
\label{sec:summary_lowmass}

In a preliminary study, and using a different subsample of the {\it Herschel} {\it DEBRIS} survey, Wyatt et al. (\citeyear{2012MNRAS.424.1206W}) identified a tentative correlation between debris and the presence of planets with masses $<$ 95 M$_\oplus$. Using a different subsample, Marshall et al. (\citeyear{2014A&A...565A..15M}) also found evidence that  stars with planets $<$ 30 M$_\oplus$ are more likely to harbor debris disks than are stars with planets $>$ 30 M$_\oplus$ (6/11 vs. 5/26). There are aspects related to the dynamical evolution of planetary systems that could result in a higher frequency of debris disks around stars with low-mass planets compared to those with high-mass planets. Wyatt et al. (\citeyear{2012MNRAS.424.1206W}) discussed two alternative scenarios: (1) if the planets formed in the outer region and migrated inward, low-mass planets would have been inefficient at accreting or ejecting planetesimals, leaving them on dynamically stable orbits over longer timescales; high-mass planets would have been more efficient at ejecting planetesimals, leaving behind a depleted population of dust-producing parent bodies. (2) Alternatively, if the planets formed in situ, the timescale for the planet to eject the planetesimals is shorter in systems with high-mass planets than with low-mass planets. However, the true migration histories of the systems studied may be significantly more complicated than the story portrayed under the two scenarios described above. For example, in our own Solar system, it is now well established that the ice giants, Uranus and Neptune, migrated outward over a significant distance to reach their current locations, sculpting the trans-Neptunian population as they did so (Hahn \& Malhotra \citeyear{2005AJ....130.2392H}). 

In this paper we have used the cleanest possible sample of the {\it Herschel} {\it DEBRIS} and {\it DUNES} surveys to assess if the data at hand can confirm the tentative detection of a low-mass planet-debris correlation. Contrary to the preliminary analyses mentioned above, here we have discarded stars without known ages, with ages $<$ 1 Gyr and with binary companions $<$100 AU, allowing us to rule out possible correlations due to effects other than planet presence. We find that the data do not show clear evidence that debris disks are more common around stars harboring low-mass planets compared to the average population. However, having a clean sample comes at a price because the smaller sample size limits the strength of the statistical result: a positive detection of a correlation could have been detected by the Fisher exact test only if the disk frequency around low-mass planet stars were to be about four times higher than the control sample.

The planet-debris disk correlation studies can shed light on the formation and evolution of planetary systems and may perhaps help ``predict" the presence of planets around stars with certain disk characteristics. Far from being a closed issue, this correlation (or lack of) needs to be revisited. In the near future, G. Bryden et al. (2015, in preparation) will address this question using a sample that combines {\it Herschel} {\it DEBRIS}, {\it DUNES}, and {\it SKARPS} surveys, overcoming to some degree our limitations due to the small sample size. However, there are another two aspects that need to be improved upon and, with the data at hand, cannot be addressed at the moment: our ability to detect fainter disks and to detect or rule out the presence of lower-mass planets to greater distances. 

Regarding the disk detections, our knowledge of circumstellar debris is limited: we only have detections for the top 20\% of the dust distribution, assuming all stars have a remnant circumstellar disk at some level; limits closer to the KB-level are only possible for nearby F+ type stars, and we are incapable of seeing exact analogues to our own Solar system leaving a large parameter space with no constraint on planet or dust properties.  Future missions under consideration such as {\it SPICA} would improve things significantly: if its telescope is not descoped, the improvement in sensitivity would allow detection of photospheres not detected by {\it Herschel}, e.g. for M stars and for FGK stars at large distances; its noise would also be lower than {\it Herschel}, allowing it to detect fainter disks.

Regarding the planet detection, the high frequency of low-mass planets indicates that we probably have many low-mass planet stars in the control sample which have not been identified, hindering our ability to detect a correlation. To overcome this problem, we rely on radial velocity surveys to gradually probe both to greater distances and lower planet masses; but also critically important is that these teams make the nondetections publicly available so we can identify systems for which the presence of planets of a given mass can be excluded out to a certain distance.

\subsection {Planetary system multiplicity}
 Dynamical simulations by Raymond et al. (\citeyear{2011A&A...530A..62R}, \citeyear{2012A&A...541A..11R}) of multiple-planet systems with outer planetesimal belts indicate that there might be a correlation between the presence of multiple planets and debris. This is because the presence of the former indicates a dynamically stable environment where dust-producing planetesimals may have survived for extended periods of time (as opposed to single-planet systems that in the past may have experienced gravitational scattering events that resulted in the ejection of other planets and dust-producing planetesimals).  It is of interest therefore to assess whether debris disks are correlated with planet multiplicity. 

Our sample does not show evidence that debris disks are more or less common, or more or less dusty, around stars harboring multiple-planet systems compared to single-planet systems.  
 
\subsection {Dust temperature} Based on the limited statistics, there is no evidence that the characteristic dust temperature of the debris disks around planet-bearing stars is any different from that in debris disks without identified planets. This is of course subject to detailed individual modeling, as the spatial dust disk distribution of the planet-bearing systems might show more structural features due to gravitational perturbations compared to the disks around stars not harboring planets, in which case it might not be appropriate to describe the dust excess emission with a single temperature.

\subsection {Stellar metallicity}
\label{sec:summary_met}
We find that there is no evidence that debris disks are more common around stars with high metallicities.  This is in agreement with previous studies (Greaves et al. \citeyear{2006MNRAS.366..283G}; Bryden et al. \citeyear{2006ApJ...636.1098B}). We find a dearth of debris disks with high dust flux ratios (also fractional luminosities) around low-metallicity stars, consistent with the model of Wyatt, Clarke \& Greaves (\citeyear{2007MNRAS.380.1737W}). However, survival analysis tests indicate that this trend is not statistically significant and that we cannot rule out the hypothesis that the high-metallicity and low-metallicity samples have been drawn from the same distribution in terms of the dust flux ratio. 

The data confirm the well-known correlation between high metallicities and the presence of high-mass planets. On the contrary, we find no evidence of a correlation between high metallicities and the presence of low-mass planets. We therefore find the well-known positive correlation between the presence of planets and stellar metallicity for stars with high-mass planets but no correlation for stars with low-mass planets only in agreement with extensive Doppler studies (Santos et al. \citeyear{2004A&A...425.1013S}; Fisher \& Valenti \citeyear{2005ApJ...622.1102F}; Ghezzi et al. \citeyear{2010ApJ...720.1290G}; Mayor et al. \citeyear{2011arXiv1109.2497M}). Maldonado et al. (\citeyear{2012A&A...541A..40M}) studied a larger stellar sample and derived the metallicities in a uniform way. They found an increasing correlation with stellar metallicity from stars without planets and disks and stars with debris disks to stars with high-mass planets. They also concluded that the correlation with stellar metallicity is due to the presence of planets and not the presence of debris disks. 

\subsection{Fractional luminosity and comparison to the Solar system debris disk}
\label{sec:summary_ss}
Comparing the observed cumulative distribution of fractional luminosity to those expected from a Gaussian distribution in logarithmic scale, we find that a distribution centered on the Solar system value (taken as 10$^{-6.5}$) fits the data well, whereas one centered at 10 times the Solar system's debris disks can be rejected. 

This is of interest in the context of future prospects for terrestrial planet detection. Even though the Herschel observations presented in this study trace cold dust located at tens of AU from the star, for systems with dust at the Solar system level, the dust dynamics is dominated by Poynting-Robertson drag. This force makes the dust in the outer system drift into the terrestrial-planet region. This warm dust can impede the future detection of terrestrial planets due to the contaminant exozodiacal emission, with its median level, its uncertainty, and shape of its distribution being some of the parameters that may affect the aperture size required for a telescope such as {\it ATLAST} to be able to characterize biosignatures (see, e.g., Stark et al. \citeyear{2014ApJ...795..122S}; Brown \citeyear{2015ApJ...799...87B}). Ruling out a distribution of fractional luminosities centered at 10 times the Solar system level implies that there are a large number of debris disk systems with dust levels in the KB region low enough not to become a significant source of contaminant exozodiacal emission.  Comets and asteroids located closer to the star are other sources of dust that can contribute to the exozodiacal emission (and for those, {\it Herschel} observations do not provide constraints), but planetary systems with low KB dust-type of emission likely imply low-populated outer belts leading to low cometary activity.  These results, therefore, indicate that there are good prospects for finding a large number of debris disk systems (i.e. systems with evidence of harboring planetesimals) with exozodiacal emission low enough to be appropriate targets for terrestrial planet searches. Dedicated warm dust surveys with the Keck Interferometer Nuller (Millan-Gabet et al. \citeyear{2011ApJ...734...67M}), CHARA/FLUOR (Absil et al. \citeyear{2013A&A...555A.104A}), VLTI/PIONIER (Ertel et al. \citeyear{2014A&A...570A.128E}) and LBTI (under the HOSTS program) are shedding or will soon shed light on this issue.

Even though the planetesimals detected by Herschel in the far infrared are located far from the terrestrial-planet region, their presence is favorable to the growth and survival of terrestrial planets because these planetesimals indicate that the system has experienced a calm dynamical evolution, as opposed to an environment of dynamically active, high-mass planets. Such an environment would tend to destroy both the outer, dust-producing planetesimal belt and the planetesimals that might otherwise build the terrestrial planets. This conclusion was the result of Raymond et al.'s (\citeyear{2011A&A...530A..62R}, \citeyear{2012A&A...541A..11R}) extensive dynamical simulations consisting of high-mass planet embryos and inner and outer belts of planetesimals. These simulations find that there is a strong correlation between the presence of cold dust in the outer planetary system and the presence of terrestrial planets in the inner region, so a system with low levels of KB dust emission might also imply a dynamical history not amicable to terrestrial planets. The Solar system, in this case, would be an outlier, with a low-level of KB dust but a high number of terrestrial planets. It would be of great interest to extend Raymond et al.'s (\citeyear{2011A&A...530A..62R}, \citeyear{2012A&A...541A..11R})  simulations to cover a wider range of initial conditions to further explore this correlation, as it would enlighten the target selection for an ATLAST-type mission. 

\begin{acknowledgements}
A.M.-M. thanks Ewan Cameron for insightful comments and the STScI DirectorÕs Discretionary Research fund for support.   
J.P.M. is a UNSW Vice Chancellor's Postdoctoral Research Fellow.
G.K and M.C.W. work was supported by the European Union through ERC grant number 279973.
C.E. and B.M. are partially supported by Spanish Ministry of Economy grant AYA 2011-26202.
D.R.R. acknowledges support from Chilean FONDECYT grant 3130520.
M.B. acknowledges support from a FONDECYT Postdoctral Fellowship, project no. 3140479.
\end{acknowledgements}

\bibliographystyle{apj} 
\bibliography{bibliography_ama} 

\clearpage

\renewcommand\thetable{2}
\LongTables
\begin{deluxetable*}{lccccccccccccc}
\tablewidth{0pt}
\tabletypesize{\scriptsize}
\tablecaption{Stellar properties}
\tablehead{
\colhead{HIP}	&	
\colhead{HD}	&	
\colhead{GJ}	&	
\colhead{UNS}	&	
\colhead{Survey}	&	
\colhead{d}	&	
\colhead{SpT}	&	
\colhead{T$_{\rm eff}$}	&	
\colhead{[Fe/H]}	&	
\colhead{L$_{\rm bol}$}	&
\colhead{Age} &
\colhead{Binary Sep.\tablenotemark{a}}	\\	
\colhead{}	&	
\colhead{}	&	
\colhead{}	&	
\colhead{}	&	
\colhead{}	&	
\colhead{(pc)}	&	
\colhead{}	&	
\colhead{(K)}	&	
\colhead{(dex)}	&	
\colhead{(L$_{\odot}$)}	&	
\colhead{(Gyr)} &
\colhead{(AU)}
}
\startdata
544	  &  166        &  5	        &  G030A       &  DUNES         &  13.68     &  G8 V                   &  5512  	  &  0.15	          &  0.631        &  0.17       &  		     \\ 	
910	  &  693        &  10	        &  F069A       &  DUNES         &  18.74     &  F8 V Fe-0.8 CH-0.5     &  6235  	  &  -0.32	          &  3.280        &  3.04       &  		     \\ 	
1368	  &             &  14	        &  K115A       &  DEBRIS        &  14.99     &  K7                     &  4089  	  &  **		          &  0.109        &  **	        &  		     \\ 	
1599	  &  1581       &  17	        &  F005A       &  DEBRIS        &  8.59      &  F9.5 V                 &  6013  	  &  -0.29	          &  1.288        &  4.03       &  		     \\ 	
3093	  &  3651       &  27	        &  K045A       &  DUNES         &  11.06     &  K0 V                   &  5261  	  &  0.2	          &  0.515        &  6.43       &  A-B: 474	     \\ 	
3583	  &  4391       &  1021         &  G041A       &  DEBRIS        &  15.17     &  G5 V Fe-0.8            &  5845  	  &  -0.2	          &  0.974        &  0.84       &  A-B: 252	     \\ 	
     	  &             &  146.01       &  G041B       &  DEBRIS        &  15.17     &  G5 V Fe-0.8            &  4228  	  &  **		          &  0.019        &  0.84       &  A-B: 252	     \\ 	
3765	  &  4628       &  33	        &  K016A       &  DEBRIS        &  7.45      &  K2.5 V                 &  5006  	  &  -0.24	          &  0.293        &  5.31       &  		     \\
\enddata	                                                                                                                                                                                                                                                                      
\tablenotetext{a}{Projected binary separation from Rodriguez et al. in preparation.} 
\label{tab:starinfo}
\end{deluxetable*}		
	
\clearpage

\renewcommand\thetable{3}
\LongTables
\begin{landscape}
\begin{deluxetable}{ccccccccccccc}
\tablewidth{0pt}
\tabletypesize{\scriptsize}
\tablecaption{Observed fluxes and excess properties}
\tablehead{
\colhead{HIP}	&	
\colhead{HD}	&	
\colhead{GJ}	&	
\colhead{UNS}	&	
\colhead{F$_{\rm obs}^{\rm100}$\tablenotemark{a}}	&	
\colhead{F$_{\rm star}^{\rm100}$\tablenotemark{b}} &	
\colhead{F$_{\rm obs}^{\rm100}/\sigma_{\rm obs}^{\rm100}$\tablenotemark{c}}	&	
\colhead{F$_{\rm obs}^{\rm100}/F_{\rm star}^{\rm100}$\tablenotemark{d}} &	
\colhead{F$_{\rm dust}^{\rm100}/F_{\rm star}^{\rm100}$\tablenotemark{d}} &	
\colhead{L$_{\rm dust}$/L$_{star}$\tablenotemark{e}}	&	
\colhead{SNR$_{\rm dust}$\tablenotemark{f}}	&	
\colhead{Dust}	&	
\colhead{Set\tablenotemark{g}}\\
\colhead{}	&	
\colhead{}	&	
\colhead{}	&	
\colhead{}	&	
\colhead{(mJy)}	&	
\colhead{(mJy)} &	
\colhead{}	&	
\colhead{}	&	
\colhead{}	&	
\colhead{}	&	
\colhead{}	&	
\colhead{detec.?}	&	
\colhead{}}
\startdata
544	  &  166        &  5	        &  G030A      &  63.55	   $\pm$  4.45	  &    7.54	 $\pm$  0.13	     &  14.28	        &     8.43      &     7.43	  &    6.49E-05       & 12.58    &  Y	        & 1,2,5	       \\	
910	  &  693        &  10	        &  F069A      &  19.29	   $\pm$  2.52	  &    14.41	 $\pm$  0.33	     &  7.65	        &     1.34      &     0.34	  & $<$2.57E-06       & 1.92     &  N	        & 1,2	       \\	
1368	  &             &  14	        &  K115A      &  41.27	   $\pm$  6.55	  &    2.85	 $\pm$  0.07	     &  6.30	        &     14.46     &     13.46	  &    9.50E-05       & 5.86     &  Y	        & 1,2,5,(conf.?)	       \\	
1599	  &  1581       &  17	        &  F005A      &  34.64	   $\pm$  2.72	  &    30.07	 $\pm$  0.52	     &  12.75	        &     1.15      &     0.15	  & $<$1.40E-06       & 1.65     &  N	        & 1,2	       \\	
3093	  &  3651       &  27	        &  K045A      &  6.61	   $\pm$  1.49	  &    10.82	 $\pm$  0.21	     &  4.42	        &     0.61      &     -0.39	  & $<$1.23E-07       & -2.79    &  N	        & 1,3a,6       \\	   	
3583	  &  4391       &  1021         &  G041A      &  11.12	   $\pm$  2.25	  &    8.05	 $\pm$  0.11	     &  4.94	        &     1.38      &     0.38	  & $<$4.40E-06       & 1.36     &  N	        & 1,2	       \\	
     	  &             &  146.01       &  G041B      &  -1.70	   $\pm$  2.17	  &    0.42	 $\pm$  0.04	     &  -0.79	        &  $<$15.41     &  $<$14.41	  & $<$1.37E-04       & -0.98    &  N	        & 1,2	       \\
\enddata	                                                                                                                                                                                                                                                                      
\label{tab:obs1}
\end{deluxetable}		

\clearpage

\LongTables
\begin{deluxetable}{ccccccccccccc}
\tablewidth{0pt}
\tabletypesize{\scriptsize}
\tablecaption{Observed fluxes and excess properties (cont.)}
\tablehead{
\colhead{HIP}	&	
\colhead{HD}	&	
\colhead{GJ}	&	
\colhead{UNS}	&	
\colhead{F$_{\rm obs}^{\rm100}$\tablenotemark{a}}	&	
\colhead{F$_{\rm star}^{\rm100}$\tablenotemark{b}} &	
\colhead{F$_{\rm obs}^{\rm100}/\sigma_{\rm obs}^{\rm100}$\tablenotemark{c}}	&	
\colhead{F$_{\rm obs}^{\rm100}/F_{\rm star}^{\rm100}$\tablenotemark{d}} &	
\colhead{F$_{\rm dust}^{\rm100}/F_{\rm star}^{\rm100}$\tablenotemark{d}} &	
\colhead{L$_{\rm dust}$/L$_{star}$\tablenotemark{e}}	&	
\colhead{SNR$_{\rm dust}$\tablenotemark{f}}	&	
\colhead{Dust}	&	
\colhead{Set\tablenotemark{g}}\\
\colhead{}	&	
\colhead{}	&	
\colhead{}	&	
\colhead{}	&	
\colhead{(mJy)}	&	
\colhead{(mJy)} &	
\colhead{}	&	
\colhead{}	&	
\colhead{}	&	
\colhead{}	&	
\colhead{}	&	
\colhead{detec.?}	&	
\colhead{}}
\startdata
29568	  &  43162      &  3389         &  G056A      &  6.57	   $\pm$  1.13	  &    5.50	 $\pm$  0.14	     &  5.81	        &     1.20      &     0.20	  & $<$3.37E-06       & 0.94     &  N	        & 1,2	       \\	
32439	  &  46588      &  240.1        &  F056A      &  8.49	   $\pm$  1.57	  &    8.58	 $\pm$  0.15	     &  5.40	        &     0.99      &     -0.01	  & $<$1.60E-06       & -0.06    &  N	        & 1,2	       \\	
32480	  &  48682      &  245          &  F044A      &  263.67	   $\pm$  20.20	  &    11.16	 $\pm$  0.45	     &  13.05	        &     23.62     &     22.62	  &    6.24E-05       & 12.49    &  Y	        & 1,2,5	       \\	
33277	  &  50692      &  252          &  G065A      &  6.66	   $\pm$  1.29	  &    7.75	 $\pm$  0.14	     &  5.15	        &     0.86      &     -0.14	  & $<$1.25E-06       & -0.84    &  N	        & 1,2	       \\	
34017	  &  52711      &  262          &  G093A      &  7.30	   $\pm$  1.25	  &    6.56	 $\pm$  0.19	     &  5.85	        &     1.11      &     0.11	  & $<$2.39E-06       & 0.59     &  N	        & 1,2	       \\	
34052	  &  53680      &  264          &  G059C      &  -0.01	   $\pm$  2.90	  &    2.42	 $\pm$  0.05	     &  0.00	        &  $<$3.60      &  $<$2.60	  & $<$2.33E-05       & -0.84    &  N	        & 1,2	       \\
34065	  &  53705      &  264.1A       &  G059A      &  8.42	   $\pm$  2.65	  &    10.01	 $\pm$  0.26	     &  3.18	        &     0.84      &     -0.16	  & $<$2.31E-06       & -0.60    &  N	        & 1,2	       \\	
34069	  &  53706      &  264.1B       &  G059B      &  1.22	   $\pm$  1.92	  &    4.28	 $\pm$  0.09	     &  0.64	        &  $<$1.63      &  $<$0.63	  & $<$3.03E-06       & -1.59    &  N	        & 1,2	       \\
\enddata	                                                                                                                                                                                                                                                                      
\label{tab:obs2}
\end{deluxetable}		

\normalsize

\clearpage

\LongTables
\begin{deluxetable}{ccccccccccccc}
\tablewidth{0pt}
\tabletypesize{\scriptsize}
\tablecaption{Observed fluxes and excess properties (cont.)}
\tablehead{
\colhead{HIP}	&	
\colhead{HD}	&	
\colhead{GJ}	&	
\colhead{UNS}	&	
\colhead{F$_{\rm obs}^{\rm100}$\tablenotemark{a}}	&	
\colhead{F$_{\rm star}^{\rm100}$\tablenotemark{b}} &	
\colhead{F$_{\rm obs}^{\rm100}/\sigma_{\rm obs}^{\rm100}$\tablenotemark{c}}	&	
\colhead{F$_{\rm obs}^{\rm100}/F_{\rm star}^{\rm100}$\tablenotemark{d}} &	
\colhead{F$_{\rm dust}^{\rm100}/F_{\rm star}^{\rm100}$\tablenotemark{d}} &	
\colhead{L$_{\rm dust}$/L$_{star}$\tablenotemark{e}}	&	
\colhead{SNR$_{\rm dust}$\tablenotemark{f}}	&	
\colhead{Dust}	&	
\colhead{Set\tablenotemark{g}}\\
\colhead{}	&	
\colhead{}	&	
\colhead{}	&	
\colhead{}	&	
\colhead{(mJy)}	&	
\colhead{(mJy)} &	
\colhead{}	&	
\colhead{}	&	
\colhead{}	&	
\colhead{}	&	
\colhead{}	&	
\colhead{detec.?}	&	
\colhead{}}
\startdata
64792	  &  115383     &  504          &  G073A      &  13.15	   $\pm$  1.24	  &    12.03	 $\pm$  0.25	     &  10.63	        &     1.09      &     0.09	  & $<$1.31E-06       & 0.89     &  N	        & 1,3a,6       \\	   	
64924	  &  115617     &  506          &  G008A      &  152.48	   $\pm$  10.03	  &    24.48	 $\pm$  0.47	     &  15.20	        &     6.23      &     5.23	  &    2.85E-05       & 12.75    &  Y	        & 1,4,5,7      \\	   	
65026	  &  115953     &  508          &  	      &  9.30	   $\pm$  1.00	  &    9.38	 $\pm$  0.26	     &  9.31	        &     0.99      &     -0.01	  & $<$4.21E-06       & -0.08    &  N	        & 1,2	       \\
65352	  &  116442     &  3781A        &  G050A      &  4.21	   $\pm$  2.19	  &    3.80	 $\pm$  0.06	     &  1.92	        &  $<$2.84      &  $<$1.84	  & $<$9.15E-06       & 0.19     &  N	        & 1,2	       \\	
65355	  &  116443     &  3782B        &  G050B      &  5.34	   $\pm$  2.20	  &    3.48	 $\pm$  0.06	     &  2.43	        &  $<$3.44      &  $<$2.44	  & $<$1.37E-05       & 0.85     &  N	        & 1,2	       \\	
65721	  &  117176     &  512.1        &  	      &  41.04	   $\pm$  2.55	  &    20.77	 $\pm$  0.42	     &  16.10	        &     1.98      &     0.98	  &    1.00E-05       & 7.85     &  Y	        & 1,3a,5,6     \\
66459	  &             &  519          &  K041A      &  1.26	   $\pm$  1.63	  &    3.33	 $\pm$  0.08	     &  0.77	        &  $<$1.85      &  $<$0.85	  & $<$9.60E-06       & -1.27    &  N	        & 1,2	       \\	
66675	  &  118926     &  521.1        &  K109A      &  6.03	   $\pm$  2.90	  &    2.09	 $\pm$  0.06	     &  2.08	        &  $<$7.04      &  $<$6.04	  & $<$7.42E-05       & 1.36     &  N	        & 1,2	       \\
\enddata	                                                                                                                                                                                                                                                                      
\label{tab:obs3}
\end{deluxetable}		

\normalsize

\clearpage

\LongTables
\begin{deluxetable}{ccccccccccccc}
\tablewidth{0pt}
\tabletypesize{\scriptsize}
\tablecaption{Observed fluxes and excess properties (cont.)}
\tablehead{
\colhead{HIP}	&	
\colhead{HD}	&	
\colhead{GJ}	&	
\colhead{UNS}	&	
\colhead{F$_{\rm obs}^{\rm100}$\tablenotemark{a}}	&	
\colhead{F$_{\rm star}^{\rm100}$\tablenotemark{b}} &	
\colhead{F$_{\rm obs}^{\rm100}/\sigma_{\rm obs}^{\rm100}$\tablenotemark{c}}	&	
\colhead{F$_{\rm obs}^{\rm100}/F_{\rm star}^{\rm100}$\tablenotemark{d}} &	
\colhead{F$_{\rm dust}^{\rm100}/F_{\rm star}^{\rm100}$\tablenotemark{d}} &	
\colhead{L$_{\rm dust}$/L$_{star}$\tablenotemark{e}}	&	
\colhead{SNR$_{\rm dust}$\tablenotemark{f}}	&	
\colhead{Dust}	&	
\colhead{Set\tablenotemark{g}}\\
\colhead{}	&	
\colhead{}	&	
\colhead{}	&	
\colhead{}	&	
\colhead{(mJy)}	&	
\colhead{(mJy)} &	
\colhead{}	&	
\colhead{}	&	
\colhead{}	&	
\colhead{}	&	
\colhead{}	&	
\colhead{detec.?}	&	
\colhead{}}
\startdata
95995	  &  184467     &  762.1        &  	      &  3.64	   $\pm$  0.78	  &    6.40	 $\pm$  0.10	     &  4.64	        &     0.57      &     -0.43	  & $<$-3.60E-07      & -3.49    &  N	        & 1,2	       \\
96100	  &  185144     &  764          &  G003       &  32.10	   $\pm$  2.66	  &    31.55	 $\pm$  0.52	     &  12.08	        &     1.02      &     0.02	  & $<$1.32E-06       & 0.20     &  N	        & 1,2	       \\
97944	  &  188088     &  770          &  	      &  9.86	   $\pm$  1.05	  &    11.23	 $\pm$  0.42	     &  9.37	        &     0.88      &     -0.12	  & $<$1.11E-06       & -1.21    &  N	        & 1,2	       \\
98698	  &  190007     &  775          &  K063A      &  4.99	   $\pm$  1.89	  &    5.39	 $\pm$  0.07	     &  2.63	        &  $<$1.98      &  $<$0.98	  & $<$7.46E-06       & -0.21    &  N	        & 1,2	       \\	
98959	  &  189567     &  776          &  G077A      &  4.64	   $\pm$  0.90	  &    6.38	 $\pm$  0.12	     &  5.15	        &     0.73      &     -0.27	  & $<$5.68E-07       & -1.92    &  N	        & 1,2,(4?,6?)      \\	   	
99240	  &  190248     &  780          &  	      &  69.14	   $\pm$  2.28	  &    70.31	 $\pm$  3.59	     &  30.28	        &     0.98      &     -0.02	  & $<$3.32E-07       & -0.28    &  N	        & 1,2	       \\
99701	  &  191849     &  784          &  K012A      &  19.42	   $\pm$  2.95	  &    10.63	 $\pm$  0.43	     &  6.59	        &     1.83      &     0.83	  & $<$2.03E-05       & 2.95     &  N	        & 1,2	       \\
99825	  &  192310     &  785          &  K027A      &  10.99	   $\pm$  1.98	  &    14.45	 $\pm$  0.54	     &  5.56	        &     0.76      &     -0.24	  & $<$9.31E-07       & -1.69    &  N	        & 1,4,7	       \\
\enddata	                                                                                                                                                                                                                                                                      
\label{tab:obs4}
\end{deluxetable}		

\clearpage

\end{landscape}

\footnotesize
\noindent $^a$~Observed PACS flux at 100 $\mu$m with 1-$\sigma$ uncertainty ($\sigma_{\rm obs}^{\rm100}$).\\
$^b$~Estimated photospheric prediction at 100 $\mu$m with 1-$\sigma$ uncertainty ($\sigma_{\rm star}^{\rm100}$).\\
$^c$~Stars with significant detected emission have F$_{\rm obs}^{\rm100}/\sigma_{\rm obs}^{\rm100} > 3.$\\
$^d$~Observed flux ratio (F$_{\rm obs}^{\rm100}/F_{\rm star}^{\rm100}$) and dust excess flux ratio (F$_{\rm dust}^{\rm100}/F_{\rm star}^{\rm100}$, where F$_{\rm dust} = F_{\rm obs}^{\rm100}-F_{\rm star}^{\rm100}$). In both cases, the 3$\sigma$ upper limits (preceeded by "$<$" symbol) are given for stars without significant detected emission and are calculated assuming the observed flux is $F_{\rm obs}^{\rm100}+3\sigma_{\rm obs}^{\rm100}$, for stars with $0 < F_{\rm obs}^{\rm100}/\sigma_{\rm obs}^{\rm100} < 3$, and 3$\sigma_{\rm obs}^{\rm100}$, for stars with $F_{\rm obs}^{\rm100}/\sigma_{\rm obs}^{\rm100} < 0$.\\
$^e$~Fractional luminosity of the dust excess emission. For stars with excess detections (SNR$_{\rm dust} > 3$), the fractional luminosity is calculated following Kennedy et al. (\citeyear{2012MNRAS.426.2115K}; \citeyear{2012MNRAS.421.2264K}). For stars with excess nondetections (SNR$_{\rm dust} < 3$), the 3$\sigma$ upper limit to the fractional luminosity is calculated from ${L_{\rm dust} \over L_{star}} = \big({T_{\rm dust} \over T_{star}}\big)^4 \big({e^{x_{\rm dust}}-1 \over e^{x_{\rm star}}-1}\big){F_{\rm obs}^{\rm100}-F_{\rm star}^{\rm100} \over F_{\rm star}^{\rm100}}$ following equation (4) in Beichman et al. (\citeyear{2006ApJ...652.1674B}), and assuming the observed flux is $F_{\rm obs}^{\rm100}+3\sigma_{\rm obs}^{\rm100}$, for stars with $F_{\rm obs}^{\rm100}/\sigma_{\rm obs}^{\rm100} > 0$, and 3$\sigma_{\rm obs}^{\rm100}$, for stars with $F_{\rm obs}^{\rm100}/\sigma_{\rm obs}^{\rm100} < 0$. In this expression, $x = {h\nu \over kT}$, where $\nu$ is the frequency correponding to 100 $\mu$m, $T_{\rm star} = T_{\rm eff}$ is the effective stellar photospheric temperature and $T_{\rm dust}$ is assumed to be 50 K (as in Eiroa et al. \citeyear{2013A&A...555A..11E}).\\
$^f$~Signal-to-noise ratio of the excess emission, given by SNR$_{\rm dust} = {F_{\rm obs}^{\rm100}-F_{\rm star}^{\rm100} \over \sqrt{{\sigma_{\rm obs}^{\rm100}}^2+{\sigma_{\rm star}^{\rm100}}^2}}$.\\
$^g$~For label information, see Table \ref{tab:set}. Systems that may be subject to confusion are labeled as "(conf.?)". The "set" classification of the systems with unconfirmed planetary systems --namely, HIP 16537 (= $\epsilon$ Eri), HIP 8102 (= $\tau$ Cet) and HIP 98959-- are indicated in parenthesis.\\
$^h$~Note that this upper limit is based on the nondetection at 100 $\mu$m; this star, however, has an excess emission at 8--35 $\mu$m with an inferred fractional luminosity of L$_{\rm dust}$/L$_{star}$ = 2$\cdot10^{-4}$ (Lisse et al. \citeyear{2007ApJ...658..584L}). 

\clearpage 

\renewcommand\thetable{4}
\begin{deluxetable*}{lccccccccccc}
\tablewidth{0pt}
\tabletypesize{\scriptsize}
\tablecaption{Planetary system properties\tablenotemark{a}}
\tablehead{
\colhead{HIP}	&	
\colhead{HD}	&	
\colhead{GJ}	&	
\colhead{UNS}	&	
\colhead{Planet}	&	
\colhead{M$_{\rm pl}$sin(i)}	&	
\colhead{a}	&	
\colhead{e} &
\colhead{R$_{\rm dust}$} &
\colhead{Set} &
\colhead{Ref.\tablenotemark{b}}\\	
\colhead{}	&	
\colhead{}	&	
\colhead{}	&	
\colhead{}	&
\colhead{Name}	&	
\colhead{(M$_{\rm Jup}$)}	&	
\colhead{(AU)}	&	
\colhead{}	&	
\colhead{(AU)}	&	
\colhead{} &
\colhead{}
}
\startdata
3093	&	3651	&	27	&	K045A	&	b	&	0.229	&	0.29	&	0.60	&				&	3a,6		& (1)\\
7513	&	9826	&	61	&	F020A	&	b	&	0.669	&	0.06	&	0.01	&				&	3a,7		& (2)\\
7513	&	9826	&	61	&	F020A	&	c	&	1.919	&	0.83	&	0.22	&				&	3a,7		& (2)\\
7513	&	9826	&	61	&	F020A	&	d	&	4.116	&	2.52	&	0.27	&				&	3a,7		& (2)\\
7978	&	10647	&	3109	&	F051A	&	b	&	0.925	&	2.02	&	0.16	&	40.3	$\pm$	5.9	&	3a,6	& (3)	\\
12653	&	17051	&	108	&	F046A	&	b	&	2.047	&	0.92	&	0.14	&				&	3a,6	 & (3)	\\
15510	&	20794	&	139	&	G005A	&	b	&	0.008	&	0.12	&	0.00	&	16.5	$\pm$	7.5	&		4,7	& (4) \\
15510	&	20794	&	139	&	G005A	&	c	&	0.007	&	0.20	&	0.00	&	16.5	$\pm$	7.5	&		4,7	& (4)\\
15510	&	20794	&	139	&	G005A	&	d	&	0.015	&	0.35	&	0.00	&	16.5	$\pm$	7.5	&		4,7	& (4)\\
26394	&	39091	&	9189	&	G085A	&	b	&	10.088	&	3.35	&	0.64	&	51.3	$\pm$	30.2	&	3a,6	& (3)	\\
27887	&	40307	&	      	&	K065A	&	c	&	0.021	&	0.08	&	0.00	&				&		4,7	& (5)\\
27887	&	40307	&	      	&	K065A	&	d	&	0.028	&	0.13	&	0.00	&				&		4,7	& (5)\\
27887	&	40307	&	      	&	K065A	&	b	&	0.013	&	0.05	&	0.00	&				&		4,7	& (5)\\
40693	&	69830	&	302	&	G022A	&	c	&	0.037	&	0.19	&	0.13	&				&		4,7	& (6)\\
40693	&	69830	&	302	&	G022A	&	d	&	0.056	&	0.63	&	0.07	&				&		4,7	& (6)\\
40693	&	69830	&	302	&	G022A	&	b	&	0.032	&	0.08	&	0.10	&				&		4,7	& (6)\\
43587	&	75732	&	324A 	&	K060A	&	e	&	0.026	&	0.02	&	0.00	&				&	3a,7	& (7)	\\
43587	&	75732	&	324A 	&	K060A	&	f	&	0.173	&	0.77	&	0.32	&				&	3a,7	& (7)	\\
43587	&	75732	&	324A 	&	K060A	&	b	&	0.801	&	0.11	&	0.00	&				&	3a,7	& (7)	\\
43587	&	75732	&	324A 	&	K060A	&	d	&	3.545	&	5.47	&	0.02	&				&	3a,7	& (7)		\\
43587	&	75732	&	324A 	&	K060A	&	c	&	0.165	&	0.24	&	0.07	&				&	3a,7	& (7)		\\
53721	&	95128	&	407	&	G033A	&	c	&	0.546	&	3.57	&	0.10	&				&	3a,7	& (8)	\\
53721	&	95128	&	407	&	G033A	&	b	&	2.546	&	2.10	&	0.03	&				&	3a,7	& (8)	\\
55848	&	99492	&	429B 	&	G079B	&	b	&	0.106	&	0.12	&	0.25	&				&	3a,6	& (3)	\\
57443	&	102365	&	442A	&	G012A	&	b	&	0.051	&	0.46	&	0.34	&				&		4,6 & (9)	\\
64924	&	115617	&	506	&	G008A	&	b	&	0.016	&	0.05	&	0.12	&	15.9	$\pm$	1.5	&		4,7 & (10)	\\
64924	&	115617	&	506	&	G008A	&	c	&	0.033	&	0.22	&	0.14	&	15.9	$\pm$	1.5	&		4,7 & (10)	\\
64924	&	115617	&	506	&	G008A	&	d	&	0.072	&	0.47	&	0.35	&	15.9	$\pm$	1.5	&		4,7 & (10)	\\
65721	&	117176	&	512.1	&		&	b	&	7.461	&	0.48	&	0.40	&	14.0			&	3a,6	& (3)	\\
67275	&	120136	&	527A 	&	F036A	&	b	&	4.130	&	0.05	&	0.02	&				&	3b,6	& (11)	\\
78459	&	143761	&	606.2	&	G064A	&	b	&	1.064	&	0.23	&	0.06	&				&	3a,6	& (3)	\\
79248	&	145675	&	614	&		&	b	&	5.215	&	2.93	&	0.37	&				&	3a,6	& (12)	\\
83389	&	154345	&	651	&	G088A	&	b	&	0.957	&	4.21	&	0.04	&				&	3a,6	& (13)	\\
86796	&	160691	&	691	&	G047	&	b	&	1.746	&	1.53	&	0.13	&				&	3a,7	& (14)	\\
86796	&	160691	&	691	&	G047	&	e	&	0.543	&	0.94	&	0.07	&				&	3a,7	& (14)	\\
86796	&	160691	&	691	&	G047	&	c	&	1.889	&	5.34	&	0.10	&				&	3a,7	& (14)	\\
86796	&	160691	&	691	&	G047	&	d	&	0.035	&	0.09	&	0.17	&				&	3a,7	& (14)	\\
99825	&	192310	&	785	&	K027A	&	c	&	0.074	&	1.18	&	0.32	&				&		4,7	& (4) \\
99825	&	192310	&	785	&	K027A	&	b	&	0.053	&	0.32	&	0.13	&				&		4,7	& (4)\\
113421	&	217107	&	      	&	G102AB	&	c	&	2.615	&	5.33	&	0.52	&				&	3a,7	& (2)	\\
113421	&	217107	&	      	&	G102AB	&	b	&	1.401	&	0.08	&	0.13	&				&	3a,7	& (2)		\\
\\
\hline
\\
\multicolumn{3}{c}{Unconfirmed planetary systems\tablenotemark{c}:}  \\
\\
16537	&	22049	&	144	&	K001A	&	b	&	1.054		&	3.38		&	0.25	&	36.0			& 3a,6 	& (15)\\
8102		&	10700	&	71	& 	G002A	& 	b	&  	0.0063		&	0.105	&	0.16 & 	8.5 			& 4,7	 	& (16) \\
8102		&	10700	&	71	& 	G002A	& 	c	&  	0.0097		&	0.195	&	0.03 & 	8.5 			& 4,7	 	& (16) \\
8102		&	10700	&	71	& 	G002A	& 	d	&  	0.011		&	0.374	&	0.08 & 	8.5 			& 4,7		& (16) \\
8102		&	10700	&	71	& 	G002A	& 	e	&  	0.013		&	0.552	&	0.05 & 	8.5 			& 4,7	 	& (16) \\
8102		&	10700	&	71	&	G002A	& 	f	&  	0.02			&	1.35		&	0.03 & 	8.5 			& 4,7	 	& (16) \\
98959	&  	189567    	& 	776  &  	G077A      & 	b	&	0.0316 		& 	0.11  	&	0.23	& 				& 3a,6 	& (17) \\
\enddata	                                                                                                                                                                                                                                                                      
\tablenotetext{a}{Planetary system properties from http://exoplanets.org.} 
\tablenotetext{b}{Orbit references are: 
(1) Wittenmyer et al. (\citeyear{2009ApJS..182...97W});
(2) Wright et al. (\citeyear{2009ApJ...693.1084W});
(3) Butler et al. (\citeyear{2006ApJ...646..505B});
(4) Pepe et al. (\citeyear{2011A&A...534A..58P}); 
(5) Mayor et al. (\citeyear{2009A&A...493..639M});
(6) Lovis et al. (\citeyear{2006Natur.441..305L});
(7) Endl et al. (\citeyear{2012ApJ...759...19E});
(8) Gregory et al. (\citeyear{2010MNRAS.403..731G});
(9) Tinney et al. (\citeyear{2011ApJ...732...31T});
(10) Vogt et al. (\citeyear{2010ApJ...708.1366V});
(11) Brogi et al. (\citeyear{2012Natur.486..502B});
(12) Wittenmyer et al. (\citeyear{2007ApJ...654..625W});
(13) Wright et al. (\citeyear{2008ApJ...683L..63W});
(14) Pepe et al. (\citeyear{2007A&A...462..769P});
(15) Hatzes et al. (\citeyear{2000ApJ...544L.145H});  
(16) Tuomi et al. (\citeyear{2013A&A...551A..79T}); 
(17) Mayor et al. (\citeyear{2011arXiv1109.2497M})
}
\tablenotetext{c}{Unconfirmed planetary systems are HD 22049 ($\epsilon$ Eri), HD 10700 ($\tau$ Cet) and HD 189567.}
\label{tab:planetprop}
\end{deluxetable*}		

\clearpage

\normalsize

\renewcommand\thetable{8}
\LongTables
\begin{landscape}
\begin{deluxetable}{lccccccccccc}
\tablewidth{0pc}
\tabletypesize{\scriptsize}
\tablecaption{Results from the statistical tests}
\tablehead{
\colhead{} &
\colhead{Unconf.} &
\colhead{Variable} &
\colhead{Set A\tablenotemark{b}} &
\colhead{Set B\tablenotemark{b}} &
\colhead{Gehan\tablenotemark{c}} &
\colhead{Log-} &
\colhead{Peto-} &
\colhead{K-S\tablenotemark{d}	} &
\colhead{Fischer's\tablenotemark{e}} &
\colhead{Poisson\tablenotemark{f}} &
\colhead{Binomial\tablenotemark{g}}\\
\colhead{} &
\colhead{planetary} &
\colhead{} &
\colhead{N$_A^{\rm tot}$(N$_A^{\rm upl}$)} &
\colhead{N$_A^{\rm tot}$(N$_A^{\rm upl}$)} &
\colhead{} &
\colhead{rank\tablenotemark{c}} &
\colhead{Prentice\tablenotemark{c}} &
\colhead{test} &
\colhead{Exact} &
\colhead{Dist.} &
\colhead{Dist.}\\
\colhead{} &
\colhead{systems\tablenotemark{a}} &
\colhead{} &
\colhead{} &
\colhead{} &
\colhead{} &
\colhead{} &
\colhead{} &
\colhead{} &
\colhead{test} &
\colhead{} &
\colhead{}
}
\startdata
\\
\multicolumn{12}{c}{\bf{Effect of stellar age on disk frequency and flux ratio:}}\\
\\
$_{1}$ & N		& Disk frequency 				& Set 2o				& Set 2y		& & & & & & 0.39 & 0.39  \\
$_{2}$ & Y		& Disk frequency 				& Set 2o				& Set 2y		& & & & & & 0.25 & 0.24  \\
\\
$_{3}$ & N  & Disk presence\tablenotemark{i}	& Set 2o	 			& Set 2y	& & & & & 0.62 & &  \\
$_{4}$ & Y  & Disk presence\tablenotemark{i}	& Set 2o	 			& Set 2y	& & & & & 0.60 & &  \\
\\
$_{5}$ & N/A  	& F$_{\rm dust}$/F$_{*}$		& Set 1o 	& Set  1y				& 0.10	& 0.03	& 0.11	&  & & &  \\
	    & 	   	& 							& 146(40) 	& 48(18)				& 		& 		& 		& &  & & 	 \\
\\	    
$_{6}$ & N		& Disk frequency 				& Set 2oo				& Set 2oy		& & & & & & 0.37 & 0.36  \\
\\
$_{7}$ & N  & Disk presence\tablenotemark{i}	& Set 2oo	 			& Set 2oy	& & & & & 0.82 & &  \\
\\
$_{8}$ & N/A  	& F$_{\rm dust}$/F$_{*}$		& Set 1oo 	& Set  1oy				& 0.20	& 0.11	& 0.12	&  & & &  \\
	    & 	   	& 							& 71(18) 	& 121(39)				& 		& 		& 		& &  & & 	 \\
	    
\hline
\\
\multicolumn{12}{c}{\bf{Effect of high-mass planet presence on disk frequency and flux ratio:}}\\
\\
$_{9}$ & N		& Disk frequency 			& Set 2o				& Set 3o		& & & & & & 0.25 & 0.27  \\
$_{10}$ & Y		& Disk frequency			& Set 2o				& Set 3o		& & & & & & 0.07 & 0.09  \\
\\
$_{11}$ & N  & Disk presence\tablenotemark{i}				& Set 2o	 	& Set 3o	& & & & & 0.41 & & \\
$_{12}$ & Y  & Disk presence\tablenotemark{i}				& Set 2o	 	& Set 3o	& & & & & 0.11 &  &  \\
\\
$_{13}$ & N 	& F$_{\rm dust}$/F$_{*}$		& Set  2		& Set 3		& 0.67	& 0.40	& 0.49	& & &  &  \\
	    & 	   	& 						& 182(62)		& 16(3)		& 		& 		& 		 & & & &  	 \\	    
$_{14}$ & Y 	& F$_{\rm dust}$/F$_{*}$		& Set  2		& Set 3		& 0.82	& 0.91	& 0.99	& & & &  \\
	    & 	  	& 						& 179(62)		& 17(3)		& 		& 		& 		 & & & &  	 \\
\\
$_{15}$ & N 	& F$_{\rm dust}$/F$_{*}$		& Set  2o	& Set 3o		& 0.56	& 0.59	& 0.48	& &  & & \\
	    & 	   	& 						& 126(37)		& 14(3)		& 		& 		& 		& &  & & 	 \\	    
$_{16}$ & Y 	& F$_{\rm dust}$/F$_{*}$		& Set  2o	& Set 3o		& 0.92	& 0.82	& 0.87	& &  & &  \\
	    & 	  	& 	
	    					& 123(37)		& 15(3)		& 		& 		& 		& &  & & 
\enddata
\label{tab:stat_result}
\end{deluxetable}

\clearpage

\LongTables
\begin{deluxetable}{lccccccccccc}
\tablewidth{0pc}
\tabletypesize{\scriptsize}
\tablecaption{Results from the statistical tests (cont.)}
\tablehead{
\colhead{} &
\colhead{Unconf.} &
\colhead{Variable} &
\colhead{Set A\tablenotemark{b}} &
\colhead{Set B\tablenotemark{b}} &
\colhead{Gehan\tablenotemark{c}} &
\colhead{Log-} &
\colhead{Peto-} &
\colhead{K-S\tablenotemark{d}	} &
\colhead{Fischer's\tablenotemark{e}} &
\colhead{Poisson\tablenotemark{f}} &
\colhead{Binomial\tablenotemark{g}}\\
\colhead{} &
\colhead{planetary} &
\colhead{} &
\colhead{N$_A^{\rm tot}$(N$_A^{\rm upl}$)} &
\colhead{N$_A^{\rm tot}$(N$_A^{\rm upl}$)} &
\colhead{} &
\colhead{rank\tablenotemark{c}} &
\colhead{Prentice\tablenotemark{c}} &
\colhead{test} &
\colhead{Exact} &
\colhead{Dist.} &
\colhead{Dist.}\\
\colhead{} &
\colhead{systems\tablenotemark{a}} &
\colhead{} &
\colhead{} &
\colhead{} &
\colhead{} &
\colhead{} &
\colhead{} &
\colhead{} &
\colhead{test} &
\colhead{} &
\colhead{}
}
\startdata
\\
\multicolumn{12}{c}{\bf{Effect of low-mass planet presence on disk frequency and flux ratio:}}\\
\\
$_{17}$ & N		& Disk frequency 			& Set 2o				& Set 4o		& & & & & & 0.18 & 0.18 \\
$_{18}$ & Y		& Disk frequency 			& Set 2o				& Set 4o		& & & & & & 0.06 & 0.05 \\
\\
$_{19}$ & N  & Disk presence\tablenotemark{i}				& Set 2o	 	& Set 4o	& & & & & 0.19 &  &  \\
$_{20}$ & Y  & Disk presence\tablenotemark{i}				& Set 2o	 	& Set 4o	& & & & & 0.07 &  &  \\
\\
$_{21}$ & N 	& F$_{\rm dust}$/F$_{*}$		& Set  2		& Set 4		& 0.29	& 0.31	& 0.34	&  & & &  \\
	    & 	   	& 						& 182(62)		& 6(0)		& 		& 		& 		 & & & &	 \\	    
$_{22}$ & Y 	& F$_{\rm dust}$/F$_{*}$		& Set  2		& Set 4		& 0.32	& 0.48	& 0.36	& &  & &  \\
	    & 	   	& 						& 179(62)		& 8(0)		& 		& 		& 		 & & & &	 \\
$_{23}$ & N 	& F$_{\rm dust}$/F$_{*}$		& Set  2o	& Set 4o		& 0.20	& 0.22	& 0.23	& &  & &  \\
	    & 	   	& 						& 126(37)		& 6(0)		& 		& 		& 		& &  & & 	 \\	
$_{24}$ & Y 	& F$_{\rm dust}$/F$_{*}$		& Set  2o	& Set 4o		& 0.23	& 0.38	& 0.25	& &  & & \\
	    & 	   	& 						& 123(37)		& 8(0)		& 		& 		& 		& &  & & \\
\hline
\\
\multicolumn{12}{c}{\bf{Effect of planet multiplicity on disk frequency and flux ratio:}} \\
\\
$_{25}$ & N		& Disk frequency 				& Set 6o				& Set 7o		& & & & & & 0.80 & 0.85 \\
$_{26}$ & Y		& Disk frequency 				& Set 6o				& Set 7o		& & & & & & 0.70 & 0.76 \\
\\
$_{27}$ & N  & Disk presence\tablenotemark{i}	& Set 6o	 			& Set 7o	& & & & & 1.0 & & \\
$_{28}$ & Y  & Disk presence\tablenotemark{i}	& Set 6o	 			& Set 7o	& & & & & 1.0 & & \\
\\
$_{29}$ & N 	& F$_{\rm dust}$/F$_{*}$		& Set  2o	& Set 6o		& 0.78	& 0.57	& 0.62	& &  & & \\
	    & 	   	& 						& 126(37)		& 10(2)		& 		& 		& 		& &  & &  \\	    
$_{30}$ & Y 	& F$_{\rm dust}$/F$_{*}$		& Set  2o	& Set 6o		& 0.92	& 0.63	& 0.78	& & & &   \\
	    & 	  	& 						& 123(37)		& 12(2)		& 		& 		& 		 & & & &  	 \\	    
$_{31}$ & N 	& F$_{\rm dust}$/F$_{*}$		& Set  2o	& Set 7o		& 0.57	& 0.42	& 0.57	& &  & & \\
	    & 	   	& 						& 126(37)		& 10(1)		& 		& 		& 		& &  & & 	 \\	    
$_{32}$ & Y 	& F$_{\rm dust}$/F$_{*}$		& Set  2o	& Set 7o		& 0.30	& 0.27	& 0.31	 & & & &  \\
	    & 	   	& 						& 123(37)		& 11(1)		& 		& 		& 		 & & & & 	 \\	     
$_{33}$ & N 	& F$_{\rm dust}$/F$_{*}$		& Set  6o	& Set 7o		& 0.66	& 0.22	& 0.56	& &  & &  \\
	    & 	   	& 						& 10(2)		& 10(1)		& 		& 		& 		& & & &  \\	    
$_{34}$ & Y 	& F$_{\rm dust}$/F$_{*}$		& Set  6o	& Set 7o		& 0.58	& 0.15	& 0.48	 & & & &  \\
	    & 	   	& 						& 12(2)		& 11(1)		& 		& 		& 		& &  & &\\
\hline
\\	    
\multicolumn{12}{c}{\bf{Effect of planet presence on dust temperature:}}\\
\\
$_{35}$ & N  & T$_{\rm dust}$				& Set 2t	 			& Set 3t \& Set 4t 	& & & & 0.80 &  & & \\
	      & 	& 							& 24				& 5			& & & & & &  	\\
$_{36}$ & Y  & T$_{\rm dust}$				& Set 2t	 			& Set 3t \& Set 4t 	& & & & 0.93 & & &  \\
	  	& 	& 							& 22				& 7			& & & & & & &
\enddata
\label{tab:stat_result}
\end{deluxetable}

\clearpage

\LongTables
\begin{deluxetable}{lccccccccccc}
\tablewidth{0pc}
\tabletypesize{\scriptsize}
\tablecaption{Results from the statistical tests (cont.)}
\tablehead{
\colhead{} &
\colhead{Unconf.} &
\colhead{Variable} &
\colhead{Set A\tablenotemark{b}} &
\colhead{Set B\tablenotemark{b}} &
\colhead{Gehan\tablenotemark{c}} &
\colhead{Log-} &
\colhead{Peto-} &
\colhead{K-S\tablenotemark{d}	} &
\colhead{Fischer's\tablenotemark{e}} &
\colhead{Poisson\tablenotemark{f}} &
\colhead{Binomial\tablenotemark{g}}\\
\colhead{} &
\colhead{planetary} &
\colhead{} &
\colhead{N$_A^{\rm tot}$(N$_A^{\rm upl}$)} &
\colhead{N$_A^{\rm tot}$(N$_A^{\rm upl}$)} &
\colhead{} &
\colhead{rank\tablenotemark{c}} &
\colhead{Prentice\tablenotemark{c}} &
\colhead{test} &
\colhead{Exact} &
\colhead{Dist.} &
\colhead{Dist.}\\
\colhead{} &
\colhead{systems\tablenotemark{a}} &
\colhead{} &
\colhead{} &
\colhead{} &
\colhead{} &
\colhead{} &
\colhead{} &
\colhead{} &
\colhead{test} &
\colhead{} &
\colhead{}
}
\hline
\\
\multicolumn{12}{c}{\bf{Effect of stellar metallicity on disk frequency and flux ratio:}}\\
\\
$_{37}$ & N/A	& Disk frequency 			& Set 1l				& Set 1h		& & & & & & 0.06 & 0.04 \\
\\
$_{38}$ & N/A  & [Fe/H]\tablenotemark{h}		& Set  1m-Set 5m		& Set 5m	& & 	& & & 0.28  & &  \\
$_{39}$ & N  & [Fe/H]\tablenotemark{h}		& Set  1m-Set 3m		& Set 3m	& & 	& & & 0.002  &  &  \\
$_{40}$ & N  & [Fe/H]\tablenotemark{h}		& Set  1m-Set 4m		& Set 4m	& & 	& & & 1.0  & &  \\
$_{41}$ & Y  & [Fe/H]\tablenotemark{h}		& Set  1m-Set 3m		& Set 3m	& & 	& & & 0.0008  & &  \\
$_{42}$ & Y  & [Fe/H]\tablenotemark{h}		& Set  1m-Set 4m		& Set 4m	& & 	& & & 0.47  & &  \\
\\
$_{43}$ & N  & [Fe/H]						& Set 2m 			& Set 3m 	& & & & 0.002 &  & &   \\
	      & 	& 							& 115				& 15		& & & & & & &   	\\	    
$_{44}$ & N  & [Fe/H]						& Set 2m 			& Set 4m 	& & & & 0.49 &  & &  \\
	      & 	& 							& 115				& 6			& & & & & &  & 	\\	    
$_{45}$ & Y  & [Fe/H]						& Set 2m 			& Set 3m 	& & & & 0.005 &  & & \\
	      & 	& 							& 112				& 16		& & & & & &  & 	\\	    
$_{46}$ & Y  & [Fe/H]						& Set 2m 			& Set 4m 	& & & & 0.32 &  & & \\
	      & 	& 							& 112				& 8			& & & & & & &  \\	    
$_{47}$ & N  & [Fe/H]						& Set 2m 			& Set 5m 	& & & & 0.33 &  & &   \\
	      & 	& 							& 115				& 26		& & & & & & &   	\\	    
$_{48}$ & Y  & [Fe/H]						& Set 2m 			& Set 5m 	& & & & 0.39 &  & & \\
	      & 	& 							& 112				& 26		& & & & & &  & 	\\	    
\\
$_{49}$ & N/A  	& F$_{\rm dust}$/F$_{*}$		& Set  1h	& Set 1h		& 0.42	& 0.27	& 0.44	 & & & &  \\
	    & 	   	& 						& 75(8)		& 61(5)		& 		& 		& 		& & & \\
\hline
\\
\multicolumn{12}{c}{\bf{Effect of spectral type on disk frequency and flux ratio:}}\\
\\
$_{50}$ & N/A	& Disk frequency 			& Set 1o (G)			& Set 1o (F)	& & & & & & 0.10 & 0.08  \\
$_{51}$ & N/A	& Disk frequency 			& Set 1o (G)			& Set 1o (K)	& & & & & & 0.88 & 0.90  \\
$_{52}$ & N/A	& Disk frequency 			& Set 1o (K)			& Set 1o (F)	& & & & & & 0.01 & 0.008  \\
\\
$_{53}$ & N/A  & Disk presence\tablenotemark{i}			& Set 1o (F+G)	 		& Set 1o (K)	& & & & & 0.16 &  &  \\
$_{54}$ & N/A  & Disk presence\tablenotemark{i}			& Set 1o (F)	 		& Set 1o (G+K)	& & & & & 0.09 &  & 
\enddata
\label{tab:stat_result}
\end{deluxetable}

\clearpage

\end{landscape}

\footnotesize
\noindent $^a$~Including unconfirmed planetary systems? "N" if those stars are included in the no-planet sample Set 2. "Y" if they are considered planet-hosts (i.e., they are included in Sets 3 or 4 and 6 or 7).\\
$^b$~N$_A^{\rm tot}$ and N$_B^{\rm tot}$ are the total number of stars in each set (detections and nondetections). The number in parenthesis  (N$_A^{\rm upl}$ and N$_B^{\rm upl}$) are the number of stars in each respective set with upper limits (i.e the number of stars with nondetections for which F$_{\rm obs}^{\rm100}/\sigma_{\rm obs}^{\rm100} <$ 3).\\
$^c$~Results from the univariate, nonparametric two-sample Gehan, logrank, and Peto-Prentice tests, indicating the probability that Sets A and B have been drawn from the same population in terms of the variable under consideration.\\
$^d$~K-S test probability. This is the probability that the cumulative distributions of the variable under consideration in Sets A and B differ by more than the observed value D, where D is a measure of the largest difference between the two cumulative distributions. A small probability implies that the distributions could be significantly different.\\
$^e$~Fisher exact test two-tail probability. \\
$^f$~Using Poisson statistics, this is the cumulative probability of finding x or more disk detections (where x is the number of disk detections in Set B), when the expected rate is that of Set A.\\
$^g$~Using a binomial distribution, this is the probability of finding x or more disk detections (where x is the number of disk detections observed in Set B), when the expected rate is that of Set A.\\
$^h$~The Fisher exact test is calculated by dividing the samples into two groups: a high metallicity with [Fe/H] $>$ -0.12 and a low metallicity with [Fe/H] $\leqslant$ -0.12. The result is the probability that Sets A and B are equally likely to have the same distribution of high vs. low [Fe/H].\\
$^i$~The Fisher exact test is calculated by dividing the samples into two groups: debris disks hosts, with a signal-to-noise ratio of the excess emission SNR$_{\rm dust} = {F_{\rm obs}^{\rm100}-F_{\rm star}^{\rm100} \over \sqrt{{\sigma_{\rm obs}^{\rm100}}^2+{\sigma_{\rm star}^{\rm100}}^2}} > 3$, and nondebris disks hosts, with SNR$_{\rm dust} < 3$. The result is the probability that Sets A and B are equally likely to harbor debris disks.\\
\clearpage

\normalsize

\renewcommand\thetable{9}
\begin{deluxetable}{lccccc}
\tablewidth{0pt}
\tabletypesize{\scriptsize}
\tablecaption{Debris disks properties (detected at 100$\mu$m)}
\tablehead{
\colhead{HIP}	&	
\colhead{HD}	&	
\colhead{GJ}	&	
\colhead{UNS}	&	
\colhead{T$_{\rm dust}^{\rm cold}$\tablenotemark{a}}	&	
\colhead{R$_{\rm dust}^{\rm cold}$\tablenotemark{a}} \\	
\colhead{}	&	
\colhead{}	&	
\colhead{}	&	
\colhead{}	&	
\colhead{(K)}	&	
\colhead{(AU)}
}
\startdata
544		&	166		&	5		&	G030A	&	86.2	$\pm$	2.0	&	8.3	$\pm$	0.4	\\
1368		&	    		&	14		&	K115A	&	29.0	$\pm$	3.2	&	30.5	$\pm$	6.8	\\
4148		&	5133		&	42		&	K089A	&	29.2	$\pm$	2.6	&	49.0	$\pm$	8.7	\\
5862		&	7570		&	55		&	F032A	&	73.8	$\pm$	23.8	&	19.8	$\pm$	12.8	\\
7978		&	10647	&	3109		&	F051A	&	49.1	$\pm$	3.6	&	40.3	$\pm$	5.9	\\
8102		&	10700	&	71		&	G002A	&	80.0	$\pm$		&	8.5	$\pm$		\\
15510	&	20794	&	139		&	G005A	&	61.8	$\pm$	14.1	&	16.5	$\pm$	7.5	\\
16537	&	22049	&	144		&	K001A	&	35.0	$\pm$	5.0	&	36.0	$\pm$		\\
16852	&	22484	&	147		&	F022A	&	98.0	$\pm$	7.7	&	14.4	$\pm$	2.3	\\
17420	&	23356	&	      		&	K087A	&	59.3	$\pm$	83.3	&	12.0	$\pm$	33.8	\\
17439	&	23484	&	152		&			&	41.0	$\pm$		&	29.0	$\pm$		\\
22263	&	30495	&	177		&	G029A	&	70.6	$\pm$	2.7	&	15.3	$\pm$	1.2	\\
23693	&	33262	&	189		&	F012A	&	115.0$\pm$	11.7	&	7.2	$\pm$	1.5	\\
26394	&	39091	&	9189		&	G085A	&	43.3	$\pm$	12.7	&	51.3	$\pm$	30.2	\\
28103	&	40136	&	225		&	F028		&	149.0$\pm$		&	8.4	$\pm$		\\
32480	&	48682	&	245		&	F044A	&	51.9	$\pm$	3.1	&	39.3	$\pm$	4.8	\\
42438	&	72905	&	311		&	G036A	&	87.2	$\pm$	9.5	&	10.2	$\pm$	2.2	\\
43726	&	76151	&	327		&	G068A	&	87.0	$\pm$	19.6	&	10.4	$\pm$	4.7	\\
61174	&	109085	&	471.2	&	F063A	&	37.4	$\pm$	1.9	&	124.6$\pm$	13.4	\\
62207	&	110897	&	484		&	F050A	&	53.7	$\pm$	8.3	&	28.2	$\pm$	8.8	\\
64924	&	115617	&	506		&	G008A	&	66.8	$\pm$	3.1	&	15.9	$\pm$	1.5	\\
65721	&	117176	&	512.1	&			&	100.0$\pm$		&	14.0	$\pm$		\\
71181	&	128165	&	556		&	K072A	&	42.5	$\pm$	59.7	&	21.0	$\pm$	59.1	\\
71284	&	128167	&	557		&	F039A	&	126.8$\pm$	34.1	&	9.1	$\pm$	4.9	\\
85235	&	158633	&	675		&	K062A	&	62.0	$\pm$	16.2	&	13.0	$\pm$	6.8	\\
107350	&	206860	&	836.7	&	G080A	&	86.6	$\pm$	8.7	&	11.0	$\pm$	2.2	\\
107649	&	207129	&	838		&	G053A	&	44.1	$\pm$	1.6	&	45.2	$\pm$	3.4	\\
114361	&	218511	&	1279		&	K114A	&	30.6	$\pm$	3.3	&	32.4	$\pm$	7.1	\\
116771	&	222368	&	904		&	F021A	&	51.3	$\pm$	29.1	&	55.1	$\pm$	62.5	\\
\enddata	                                                                                                                                                                                                                                                                      
\tablenotetext{a}{T$_{\rm dust}^{\rm cold}$ and R$_{\rm dust}^{\rm cold}$ for the stars with 100 $\mu$m excesses, calculated following Kennedy et al. (\citeyear{2012MNRAS.426.2115K}; \citeyear{2012MNRAS.421.2264K}) using the full spectral energy distribution.} 
\label{tab:diskprop}
\end{deluxetable}		

\end{document}